\documentclass[aps,pre,url,twocolumn 
,superscriptaddress,isbn=false]{revtex4-2}

\usepackage{latexsym}
\usepackage{amssymb,amsmath,amsbsy}
\usepackage{graphicx,color}
\usepackage{bm}
\usepackage{longtable}
\usepackage{epsf}
\usepackage[utf8]{inputenc}
\usepackage{enumerate}
\usepackage[english]{babel}               

\usepackage[colorlinks=true,
linkcolor=RoyalBlue,
citecolor=RedViolet,
urlcolor=RoyalBlue
]{hyperref}
\usepackage[dvipsnames]{xcolor}

\definecolor{amber}{rgb}{1.0, 0.49, 0.0}

\newcommand{\D}{\mathcal{D}}
\newcommand{\HH}{\mathcal{H}}

\newcommand{\Z}{\mathcal{Z}}
\newcommand{\s}{\mathcal{S}}

\newcommand{\dd}{\mathrm{d}}

\newcommand{\AAA}{\mathcal{A}}

\newcommand{\G}{\Gamma}

\newcommand{\xx}{{\bm x}}

\newcommand{\kk}{{\bm k}}
\newcommand{\pp}{{\bm p}}
\newcommand{\qq}{{\bm q}}

\newcommand{\rr}{{\bm r}}

\newcommand{\tint}{{\textstyle\int}}

\allowdisplaybreaks

\usepackage{comment}

\begin{document}


\title {Fixed $ d $ Renormalization Group Analysis of Conserved Surface Roughening}
\author{Viktor \v{S}kult\'ety}
\email{viktor.skultety@ed.ac.uk}
\affiliation{SUPA, School of Physics and Astronomy, The University of Edinburgh, Peter Guthrie Tait Road, Edinburgh EH9 3FD, United Kingdom
}
\author{Juha Honkonen}
\affiliation{Department of Military Technology, National Defence University, Santahaminantie 2, 00860 Helsinki, Finland
}

\date{\today}
\begin{abstract}

	Conserved surface roughening represents a special case of interface dynamics where the total height of the interface is conserved. Recently, it was suggested [F.  Caballero \emph{et al.},  Phys. Rev. Lett. \textbf{121},  020601 (2018)] that the original continuum model known as `Conserved Kardar-Parisi-Zhang'(CKPZ) equation is incomplete, as additional non-linearity is not forbidden by any symmetry in $ d > 1 $. 
	In this work, we perform detailed field-theoretic renormalization group (RG) analysis of a general stochastic model describing conserved surface roughening. Systematic power counting reveals additional marginal interaction at the upper critical dimension, which appears also in the context of molecular beam epitaxy. 
	Depending on the origin of the surface particle's mobility, the resulting model shows two different scaling regimes; If the particles move mainly due to the gravity, the leading dispersion law is $ \omega \sim k^{2} $, and the mean-field approximation describing a flat interface is exact in any spatial dimension. On the other hand, if the particles move mainly due to the surface curvature, the interface becomes rough with the mean-field dispersion law $ \omega \sim k^{4} $, and the corrections to scaling exponents must be taken into account. We show, that the latter model consist of two sub-class of models that are decoupled in \emph{all} orders of perturbation theory. Moreover, our RG analysis of the general model reveals that the universal scaling is described by a rougher interface than CKPZ universality class.   The universal exponents are derived within the one-loop approximation in both fixed $ d $ and $ \varepsilon $-expansion schemes, and their relation is discussed. We point out all important details behind these two schemes which are often overlooked in the literature, and their misinterpretation might lead to inconsistent results.
	
\end{abstract}

\maketitle

\section{Introduction \label{sec:intro}}

An interface is a system that is regularly found in our everyday life, starting from any material surface, or biological membrane, to flame front or landscape \cite{barabasi1995}. It is important to understand scaling properties of these structures, which will help us to identify universal laws that unify different physical phenomena.

From a theoretical point of view, the statistical properties of surfaces have been extensively studied for several decades using both analytical methods, and  computer simulations \cite{Marsili1996,Krug1997}. Based on the latter, Family and Vicsek conjectured self-similar structure for the width of a stochastically evolving interface, and extracted spatio-temporal power laws \cite{Family1985}. Later, Edwards and Wilkinson \cite{edwards1982} considered a linear stochastic differential equation for the relative interface height, and extracted a simple power law relations. Many experiments, however, show non-trivial scaling properties that cannot be explained with a linear model \cite{barabasi1995}. Hence, Kardar, Parisi and Zhang put forward a non-linear equation based on geometrical arguments, also known as KPZ equation, and predicted non-trivial scaling laws. Since then, the properties of KPZ equation and its generalizations have been extensively studied \cite{Tauber12}.

In general, equations describing interfacial dynamics can be derived phenomenologically based on symmetries~\cite{barabasi1995}, or using differential geometry on the basis of Monge Gauge \cite{Krug1997,Marsili1996}. The corresponding models are often non-linear which requires usage of more sophisticated methods. Many of the interfacial models show scale-invariant regimes, which are responsible for occurrence of fractal structures with non-trivial scaling laws. 
Probably the most effective method for studying universal properties of scale invariant field theories is the renormalization group (RG) \cite{Bogolyubov1955,Wilson1974}. In stochastic processes, the field-theoretic RG was pioneered by De Dominicis \cite{DeDominicis1976} and Janssen \cite{Janssen1976} (for Wilson's approach see \cite{Ma1975,*FNS77}), and it proved to be an excellent tool for understanding universal scaling of surface roughening \cite{Tauber12,Canet2010}.

In some interfacial systems, the total height occupied by surface particles is conserved~\cite{Racz91}, which leads to a process commonly referred to as conserved surface roughening. The theoretical analysis of field equations obeying the conservation law was first performed in the work by Sun \emph{et al} \cite{Sun1989}. Their model represents a minimal extension of the standard KPZ equation written in the form of continuity equation, also known as CKPZ equation. Recently, Cabalero \emph{et al.} \cite{Caballero2018} pointed out that additional non-linearity exists above one dimension which contains the same number of gradients, and powers in the field (four gradients and two fields). Their RG calculation revealed a run-away solution in the certain range of the parameter space, which suggests a possible strong coupling behavior. However, higher order non-linearities discarded in \cite{Caballero2018} do not necessary need to be irrelevant for the description of a perturbative fixed point, as similar scenario occurs in other interfacal models as well \cite{Haselwandter2008,Antonov1995}.

Here, we study the effect of higher order nonlinearities in the gradient expansion, containing up to four gradients and three powers in the field, as it turns out that the latter term, at the marginal dimension $ d=2 $, has the same relevance as the nonlinearities considered in \cite{Caballero2018}. However, due to the structure of the perturbation theory this term is not generated in \emph{any} order of the perturbation theory, as long as its initially absent in the model. Nevertheless, the general model shows an intriguing universal behavior which properties we derive within the one-loop approximation.

As it is standard in the perturbative RG analysis, the scaling properties in the critical region can be derived using the `$ \varepsilon $-expansion' \cite{Zinn}. The main idea is to expand the system around the Gaussian fixed point, and calculate the universal scaling exponents in a series of $ \varepsilon = d_{c} - d $, where $ d_{c} $ is the upper critical dimension. In the RG approach to surfaces however, it is very popular to use so-called `fixed $ d $' approach, where the entire calculation is performed in fixed dimension without any series expansion in $ \varepsilon $ \cite{Tauber12}. Besides that this approach is, strictly speaking, \emph{not} controlled, it requires special attention when computing Feynman diagrams, that is often overlooked in the literature. We will therefore carry out the entire calculation within both $ \varepsilon $-expansion and fixed $ d $ schemes, and discuss their mutual relation.

This manuscript is organised as follows. In Sec \ref{S2} we begin with discussing general form of Langevin equation describing conserved surface roughening. Previously analysed special cases, and corresponding critical properties are then briefly mentioned. As the following sections are rather technical, we finish Sec. \ref{S2} with a brief discussion of our results. In Sec. \ref{S3} the problem is reformulated into the De Domincis-Janssen functional integral formalism, and the structure of the perturbation theory is analyzed. As the details about the ultraviolet (UV) renormalization, and its relation to critical properties of stochastic field theories are mostly absent in the literature, in Sec \ref{S3:} we discuss this topic in detail. In section \ref{S4}, the infrared (IR) fixed points and critical exponents are derived in the context of both $ \varepsilon $-expansion and fixed $ d $ scheme. At the end we compare our results with previously obtained results. Details of the calculation can be found in the Appendices \ref{APP:DoD}, \ref{APP:FD}, and \ref{APP:CC}.

\section{Conserved surface roughening} \label{S2}

\subsection{General formulation}

We are interested in the dynamics of a surface/interface, where the total volume is conserved. Hence, the continuum description is given in terms of the height field $ \phi(\xx,t) $, with $ \xx $ being $ d $-dimensional position vector, which must obey the conservation law $ \partial_{t} \int \dd \xx \ \phi = 0 $. The physically relevant spatial dimension is either $ d = 1 $ or $ 2 $, depending whether the system is confined to a plane or not. The conservation law implies that the dynamical equation must have a form of the continuity equation~\cite{barabasi1995,Marsili1996,Krug1997}
\begin{align}
\partial_{t} \phi = - \partial_{i} ( j_{i}^{D} +  j_{i}^{R} ), \label{S2:B.ContEq}
\end{align}
where the summation over the repeated indices is implied. In \eqref{S2:B.ContEq}, $ j_{i}^{D} $ is the `deterministic' current of the surface particles due to the interface roughness, and $ j_{i}^{R} $ is a `random' current responsible for stochastic movement of the surface particles due to the thermal fluctuations. Naturally, $ j_{i}^{D} $ can depend only on the surface derivatives $ \partial \phi $, and the random force $ f \equiv  - \partial_{i} j_{i}^{R} $ obeys Gaussian statistics with zero mean and two-point correlator
\begin{align}
\langle f(\xx,t) f(\xx',t') \rangle &= - \AAA \partial^{2} \delta(t-t') \delta(\xx-\xx'). \label{S2:B.noise}
\end{align}
In \eqref{S2:B.noise}, $ \AAA $ is the noise amplitude, whose exact form depends on the scaling regime under consideration (see below). Note that Eq. \eqref{S2:B.ContEq} has very similar structure to equations discussed in the context of Molecular Beam Epitaxy (MBE) \cite{barabasi1995,Marsili1996,Krug1997,Haselwandter2008}. The difference comes from the noise correlator \eqref{S2:B.noise}, which in MBE is non-conservative due to incoming particle flux to the surface.

Knowing the solution to \eqref{S2:B.ContEq}, with  \eqref{S2:B.noise}, the quantity of interest is the interface width \cite{barabasi1995,Krug1997}
\begin{align}
W(L,t) \equiv \left\langle  \frac{1}{L^{d}} \int \dd \xx \ \phi^{2}(\xx,t) \right\rangle^{1/2}, \label{S2:B.width}
\end{align}
where the average is taken over the random noise history, and $ L^{d} $ is the system size in $ d $-dimensions. Starting from a flat interface, Family and Vicsek conjectured \cite{Family1985} that the width \eqref{S2:B.width} should have the following form
\begin{align}
W(L,t) = L^{\chi} F(t/L^{z}), \label{S2:B.roughness}
\end{align}
where $ \chi $ and $ z $ are scaling exponents (the latter is usually referred to as dynamical critical exponent). Depending on the ratio $ x \equiv t/L^{z} $, the expression \eqref{S2:B.roughness} shows two distinct regimes
\begin{itemize}
	\item[i)] In the long time limit and at fixed $ L^{z} \ll t $, the system is expected to saturate
	\begin{align}
	F(x \gg 1) \sim \text{const}., \ \implies \ W(L,t) \sim L^{\chi}.
	\end{align}
	The surface is said to be asymptotically rough for $ \chi > 0 $, and asymptotically flat for $ \chi < 0 $. Moreover, the roughness depends on the scale of observation~\cite{Krug1997}.
	
	\item[ii)] Before the saturation occurs $ 0 \ll t \ll L^{z} $, the correlations are expected to be independent of the system size
	\begin{align}
	F(x \ll 1) \sim x^{\chi/z}, \ \implies \ W(L,t) \sim t^{\chi/z}.
	\end{align}
\end{itemize}
The above exponents can be calculated from the Langevin equation \eqref{S2:B.ContEq}, by computing the correlations of the $ \phi $ field
\begin{align}
\left\langle \phi(\xx,t) \phi(\xx',t') \right\rangle =  |\xx-\xx'|^{2\chi} C \left( \frac{t-t'}{|\xx-\xx'|^{z}} \right). \label{S2:B.corr}
\end{align}

\subsection{Stochastic field equations} \label{S2:B}

The simplest dynamical equation describing conserved surface roughening of type \eqref{S2:B.ContEq} can be obtained by performing gradient expansion of the linear stochastic differential equation \cite{edwards1982,barabasi1995}
\begin{align}
\partial_{t} \phi =  ( \tau \partial^{2} - D \partial^{4} + \dots ) \phi + f, \label{S2:B.LE}
\end{align}
where $ \tau $ and $ D $ can be both viewed as diffusion constants: $ \tau $ is related to the effects of gravity on the the surface while $ D $ is related to the surface curvature. This is because they can be derived from the chemical potentials $ \mu_{\text{ch}} = -\tau\phi $, and $ D \partial^{2}\phi $, respectively. Since Eq.~\eqref{S2:B.LE} can be written in the form
\begin{align}
\partial_{t} \phi = -\partial^{2} \tfrac{\delta }{\delta \phi}\HH[\phi] + f, \label{S1:eq.H}
\end{align}	
 where $\HH$ is an effective Hamiltonian, it describes a relaxational process towards an equilibrium state given by a flat interface $ \phi = 0 $. Depending on the length scale of interest $ l $, Eq.  \eqref{S2:B.LE} contains information about two different regimes. The crossover between them is set by the ratio of diffusion constants \cite{barabasi1995,Krug1997}
\begin{align}
l \ll \sqrt{\tau/D}: &\quad z = 2, \quad \chi =  -d/2 , \hspace{0.96cm} \AAA = \tau, \label{S1:eq.z2} \\
l \gg \sqrt{\tau/D}: &\quad z = 4, \quad \chi = (2-d)/2 , \quad \AAA = D, \label{S1:eq.z4}
\end{align}
where the amplitude of the noise is rescaled with the corresponding amplitude of the leading linear term for the convinience. 

Once the nonlinearities are considered, the results \eqref{S1:eq.z2} and \eqref{S1:eq.z4} can be also interpreted differently; in the context of phase transitions, $\tau$ plays similar role of the deviation from the criticality, which value determines the leading dynamical scaling behavior. If $ \tau \gg 1 $, which occurs when the movement of the particles is caused mainly by the graviational effects, all higher order terms in \eqref{S2:B.ContEq} are irrelevant from the RG point of view, and the scaling \eqref{S1:eq.z2} is exact in any realistic spatial dimension. We will justify this statement in Sec. \ref{S6:D}. On the other hand, if $ \tau \ll 1 $, which occurs when the movement of particles is mainly due to the surface curvature, the system is found to be in a different critical regime with the mean-field scaling properties \eqref{S1:eq.z4}, and where the corrections coming from nonlinearities must be taken into account using a RG procedure. The analysis of the latter regime represents the main purpose of this work.

The first non-linear stochastic field equation \eqref{S2:B.ContEq} studied in the context of conserved surface roughening was introduced by Sun \emph{et al.} \cite{Sun1989}. Authors postulated that the relevant equation can be found by writing the original KPZ equation in a conservative form \eqref{S2:B.ContEq}
\begin{align}
\partial_{t} \phi = -\partial^{2} \left[ D \partial^{2} \phi + \tfrac{1}{2} \lambda (\partial_{i}\phi)^{2} \right] + f. \label{S2:B.CKPZ86}
\end{align}
This model, usually referred to as conserved KPZ equation (CKPZ), is purely of a non-equilibrium nature as it cannot be written in a form \eqref{S1:eq.H}. The Wilson's RG analysis performed by Sun \emph{et al.} revealed that systems described by the CKPZ equation belong to a new universality class with non-trivial critical exponents
\begin{align}
z = 4 -  \tfrac{1}{3} \varepsilon + \mathcal{O}(\varepsilon^{2}), \quad \chi = \tfrac{1}{3} \varepsilon + \mathcal{O}(\varepsilon^{2}),  \label{S2:B.CKPZ86_exponents}
\end{align}
where $ \varepsilon = 2 - d $ is the deviation from the upper critical dimension $ d_{c} = 2 $. The above exponents seem to satisfy the relation $ z + \chi = 4 $, analogous to $ z + \chi = 2 $ in non-conserved KPZ equation \cite{KPZ86}. The relation holds due to the absence of the renormalization of the quadratic non-linearity (the coupling constant is renormalized), which in KPZ equation is the result of Galilean invariance. In analogy to non-conserved KPZ equation, Sun \emph{et al.} proposed a suspicious analogue of Galilean invariance which forbids the renormalization of the non-linearity in all orders of perturbation theory. Janssen showed however~\cite{Janssen1997}, that the two-loop contributions do exist but are too small for experimental verification.

Caballero \emph{et al.} \cite{Caballero2018} recently proposed a modification of the original CKPZ equation \eqref{S2:B.CKPZ86}; they suggested, that in $ d > 1 $ the full stochastic field equation for the conserved surface roughening may have the following form
\begin{align}
\partial_{t} \phi =&\ - D \partial^{4} \phi -\tfrac{1}{2} \lambda \partial^{2} (\partial \phi)^{2} \nonumber \\
&\hspace{1.42cm} + \tfrac{1}{2} \zeta \partial_{i} ((\partial_{i}\phi)(\partial^{2}\phi)) + f. \label{S1:eq.CKPZplus}
\end{align}
This model is referred to as CKPZ+ equation.  The extra $ \zeta $ non-linearity (which coincides with $ \lambda $ non-linearity in $ d = 1 $) is not forbidden by any symmetry, and it also cannot be written in the form \eqref{S1:eq.H}. Although the perturbative fixed point revealed by the RG analysis of \eqref{S1:eq.CKPZplus} appeared to have exactly CKPZ critical exponents \eqref{S2:B.CKPZ86_exponents}, a runaway solution possibly describing a strong coupling regime was also reported. We would like to point out that in the limit $ \zeta = 2 \lambda $ the non-linearities in \eqref{S1:eq.CKPZplus} can be derived from an effective Hamiltonian in a context of non-conservative dynamics \cite{Escudero08}. Note that such non-linearity vanishes in $ d = 1 $.

The last model we would like to mention here is
\begin{align}
\partial_{t} \phi = (\tau \partial^{2}- D \partial^{4}) \phi + \tfrac{1}{3!} \eta \partial_{i} ((\partial_{i}\phi)(\partial_{j}\phi)^{2}) + f. \label{S1:eq.LE-IR}
\end{align}
This equation first appeared in \cite{Lai1991} in the context of MBE, where it was suggested that the cubic non-linearity represents corrections to the $ \tau\partial^{2}\phi $ term. Although the effect of this non-linearity on the leading IR behavior of MBE has been analysed \cite{SarmaKotlyar94,Haselwandter2008}, its influence on the conserved surface roughening (with CKPZ+ term) is, to the best of our knowledge, unknown.

To summarize--in what follows, we will be interested in analysing critical scaling of the general equation describing conserved surface roughening
\begin{align}
\partial_{t} \phi =&\ (\tau \partial^{2} - D \partial^{4}) \phi  -\tfrac{1}{2} \lambda \partial^{2} (\partial \phi)^{2} + \tfrac{1}{2} \zeta \partial_{i} ((\partial_{i}\phi)(\partial^{2}\phi)) \nonumber \\
&\ + \tfrac{1}{3!} \eta \partial_{i} ((\partial_{i}\phi)(\partial\phi)^{2}) + f, \label{S1:eq.LE}
\end{align}
with the noise correlator \eqref{S2:B.noise}. Detailed analysis of \eqref{S1:eq.LE} in the context of MBE (i.e. with non-conservative noise) can be found in \cite{Sherman2012,Sherman2012a}.

As the following chapters are quite technical, we will briefly mention our main results at this stage. As stated above, Eq. \eqref{S1:eq.LE} contains information about two fundamentally different scaling regimes; for $ \tau/D \gg 1 $, the leading dynamical scaling is $ \omega \sim k^{2} $, where all nonlinearities are irrelevant and the exponents \eqref{S1:eq.z2} are exact at any $ d $. For $ \tau/D \ll 1 $, the leading scaling is $ \omega \sim k^{4} $, where corrections to \eqref{S1:eq.z4} coming from nonlinearities must be taken into account. Systematic power counting of the latter case reveals that all nonlinearities considered in \eqref{S1:eq.LE} are marginal at the upper critical dimension $ d_{c} = 2 $. The reason for this is the vanishing canonical dimension of the field $\phi$ at $d=d_c$, so that the `RG relevance' of any interaction is determined solely from the number of $\partial$ operators. Nevertheless, the models \eqref{S1:eq.CKPZplus} and \eqref{S1:eq.LE-IR} are decoupled in \emph{all} orders of perturbation theory; renormalization of the model \eqref{S1:eq.LE-IR} does not generate CKPZ(+) nonlinearities due to the $ \phi \leftrightarrow -\phi $ symmetry, while the opposite argument is rather technical, and related to the structure of the perturbation theory. 

Intriguingly, our analysis of the general model \eqref{S1:eq.LE} shows that for $ d < d_{c} $ the RG flow reaches a fixed point where the CKPZ(+) nonlinearities are \emph{irrelevant}, and the universal scaling is determined by the much simpler model \eqref{S1:eq.LE-IR}. There is, however, still a region in the parameter space where the RG flow shows a runaway solution. Similarly to the CKPZ+ model, this might be a sign of a strong coupling behavior. The most non-trivial fixed point with all nonlinearities being relevant also appears, but it is unstable under any circumstances. The corrections to the critical exponents \eqref{S1:eq.z4} are absent in the one-loop approximation, as it is standard for stochastic field theories with cubic interactions (the only exception is scaling of $ \tau $)~\cite{Vasilev2004,Tauber12}. Further discussion about relation of our results with previous works, and commonly overlooked technical details in Wilson's RG are found in Sec.~\ref{S4.E}.

\section{Field-theoretic formulation} \label{S3}

We will now proceed to the calculation of
critical properties of the model \eqref{S1:eq.LE} beyond mean-field approximation. The most straightforward way is to deploy Wilson's RG approach, i.e. to integrate out the fast modes, and store the corresponding contributions in the renormalized parameters and fields. Although this approach is quite intuitive, it is inconvenient for practical calculations. Instead, we shall use the field-theoretic RG which represents a more systematic approach \cite{Zinn,Vasilev2004}. The starting point is to map the model \eqref{S1:eq.LE} into the functional integral formalism, where the UV renormalization will allow us to calculate corrections to the mean-field scaling exponents. Most of the following calculations will be given in the context of $ \omega \sim k^{4} \rightarrow 0 $ limit, if not stated otherwise.

\subsection{Response functional} \label{S2:C}

Using standard methods \cite{DeDominicis1976,Bausch1976,Vasilev2004,Tauber12} we derive De Dominicis-Janssen response functional for the stochastic problem \eqref{S1:eq.LE},
\eqref{S2:B.noise}, with $\AAA\to D_0$,

\begin{align}
\s[\varphi] =&\ \s^{\text{free}}[\varphi] + \s^{\text{int}}[\varphi] \label{S3:eq.RF} \\
 \s^{\text{free}}[\varphi] =&\ \tfrac{1}{2} D_{0} \phi' \partial^{2}
 \phi' + \phi' \{ \partial_{t} +  D_{0} (\partial^{2} - \tau_{0}) \partial^{2} \} \phi \label{S3:eq.RFa} \\
\s^{\text{int}}[\varphi] = &\ D_{0} \big[ \tfrac{1}{2} \lambda_{0} (\partial^{2}\phi') (\partial\phi)^{2} + \tfrac{1}{2} \zeta_{0} ( \partial_{i} \phi' )(\partial_{i} \phi) (\partial^{2} \phi) \nonumber \\
& \hspace{0.7cm} + \tfrac{1}{3!} \eta_{0} ( \partial_{i} \phi' )(\partial_{i} \phi) (\partial \phi)^{2}  \big], \label{S3:eq.RFb}
\end{align}
where $ \varphi = \{ \phi',\phi \} $,  the subscript $ 0 $ stands for the bare parameters and we have rescaled parameters as $ g_{0} \rightarrow D_{0} g_{0}, \ g_{0} = \{ \lambda_{0},\zeta_{0},\eta_{0} \} $ due to dimensional reasons (see below). The Martin-Siggia-Rose \cite{MSR} response field $ \phi' $ is the Langrange multiplier that enforces the Langevin equation \eqref{S1:eq.LE}. The latter appears after performing the functional integral over the noise in the process of deriving the response functional  \eqref{S3:eq.RF}.  The final form of \eqref{S3:eq.RF} is also uniquely defined regardless of whether Ito or Stratonovich interpretation of the stochastic differential equation is used, as the noise \eqref{S2:B.noise} is not multiplicative~\cite{honkonen2012ito}.

In expressions such as \eqref{S3:eq.RF}, the integrations over all spatial and temporal variables is implicitly assumed. For example, the first term represents the following expression
\begin{align}
\phi'\partial^{2}\phi' \equiv \int \dd \xx \dd t \ \sum_{i=1}^{d} \ \phi'(\xx,t) \partial_{i}\partial_{i}\phi'(\xx,t). 
\end{align}
where $ d $ is the dimensionality of the space. We finalize this theoretical setup with standard assumption of the vanishing initial and boundary conditions
\begin{align}
\varphi(|\xx|,t\rightarrow-\infty) = \varphi(|\xx|\rightarrow\pm\infty,t) =  0.
\end{align}

The field-theoretic formulation \eqref{S3:eq.RF} implies, that correlations of all fields $ \phi $ and response fields $ \phi' $ can be calculated directly from the following generating functional
\begin{align}
\Z[H] = \int \D \varphi \exp\{ -\s[\varphi] + \varphi H \}, \label{S2:eq.Z}
\end{align}
where $ \varphi H = \phi h + \phi'h' $, by taking the corresponding number of variational derivatives with respect to the source fields $ H = \{ h,h' \} $, at $ H = 0 $. For example the two-point correlation function can be calculated as follows
\begin{align}
\langle \phi(\xx,t) \phi(\xx',t') \rangle = \frac{\delta^{2} \Z[H]}{\delta h(\xx,t)\delta h(\xx',t')} \bigg|_{H=0}. \label{S2:eq.ZG}
\end{align}
The response field $ \phi' $ in \eqref{S3:eq.RF} is then used for generating the response functions, for example
\begin{align}
\left\langle \frac{\delta \phi(\xx,t)}{\delta f(\xx',t')} \right\rangle \Longleftrightarrow \langle \phi(\xx,t) \phi'(\xx',t') \rangle.
\end{align}

We would like to stress out that there is in principle no difficulty of carrying out the functional integral over the response field $ \phi' $ in the expression \eqref{S2:eq.Z}, as \eqref{S3:eq.RF} is quadratic in $ \phi' $. This would lead to the generating functional whose integrand can be interpreted as a `path probability distribution' for the evolution of the field $ \phi $, with the resulting `action' functional usually referred to as Onsanger-Machlup functional \cite{Onsager1953,Machlup1953}. 
However, the field-theoretic renormalization proceduce used in our analysis is designed for response functionals local in time and space,
therefore further calculations will be carried out with the response functional \eqref{S3:eq.RF}.

\subsection{Perturbation theory} \label{S2:PT}

\begin{figure}[t!]
	\centering
	\includegraphics[width=7cm]{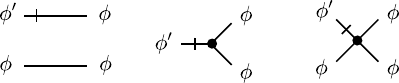}
	\caption{Graphical representation of Feynman rules.}
	\label{fig:FR}
\end{figure}

The only unambiguous and mathematically rigorous definition of the functional integral \eqref{S2:eq.Z} for the interacting field theories such as \eqref{S3:eq.RF} is through Gaussian integrals \cite{Faddeev18} and gives rise to the perturbation theory. In the present work, we will therefore perform perturbative analysis of the model. This is done by expanding the interaction part $ \s^{\text{int}}[\varphi] $ in terms of the bare coupling constant(s) $ g_{0} $, where the averages such as \eqref{S2:eq.ZG} are then evaluated with respect to the quadratic (`free') theory $ \s^{\text{free}}[\varphi] $. These calculations are schematically represented in terms of Feynman diagrams, whose graphical representation is depicted in Fig. \ref{fig:FR}: The propagators are determined from the inverse of the quadratic part $ \s^{\text{free}}[\varphi] $ of the response functional \eqref{S3:eq.RF}, and the vertex factors represent the multiplicative factors of the interaction parts $ \s^{\text{int}}[\varphi] $. Translation invariance in space and time gives rise to Fourier representations
\begin{align}
\langle \phi\phi' \rangle_{0}(k,\omega) &=
 \int \dd \xx \dd t \ \langle \phi(\xx,t) \phi'(0,0)\rangle_{0}  {\rm e}^{-i(\kk.\xx-\omega t)}
 \nonumber\\
 &=
\frac{1}{-i\omega + D_{0} k^{2}(k^{2} + \tau_{0})}, \label{S2:eq.p1} \\
\langle \phi\phi \rangle_{0}(k,\omega) &= \frac{D_{0}k^{2}}{\omega^{2} + D_{0}^{2} k^{4}(k^{2} + \tau_{0})^{2} }, \label{S2:eq.p2} \\
\mathcal{V}_{\phi_{\pp}'\phi_{\qq}\phi_{\rr}} &= - \tfrac{1}{2} D_{0} \big[ 2\lambda_{0} p^{2} (\rr.\qq) \nonumber \\
&+ \zeta_{0} ( r^{2} (\pp.\qq) +  (\rr.\pp)q^{2} ) \big]
\label{S2:eq.v1} \\
\mathcal{V}_{\phi_{\pp}'\phi_{\qq}\phi_{\rr}\phi_{\bm{o}}} &= -\tfrac{1}{3}D_{0} \eta_{0} \big[ (\pp.\qq)(\rr.\bm{o})  + (2 \ \text{perm.}) \big]. \label{S2:eq.v2}
\end{align}
with conservation laws of momentum and energy in vertices ($ \rr = -\pp - \qq $ in \eqref{S2:eq.v1} and $ \bm{o} = -\pp-\qq-\rr $ in \eqref{S2:eq.v2}).

The above rules allow us to calculate perturbative corrections to observables such as response \eqref{S2:eq.p1} and correlation functions \eqref{S2:eq.p2}. We are interested in the critical point limit, i.e. in the limit $ \omega \sim k^{z} \rightarrow 0 $, with all relevant parameters going to zero (see below). In field theories, this limit is normally singular, as the individual perturbative corrections diverge. In the proceeding section we will describe how to extract information about the critical point by renormalizing the model.

As we will be interested in renormalizing the model, it is much more convenient to work with the effective action
\begin{align}
    \G[\Psi] = \mathcal{W}[H] - H \Psi, \quad \mathcal{W}[H] = \ln \Z[H], \label{pert.theor:eq.Gamma}
\end{align}
instead of \eqref{S2:eq.Z}. In \eqref{pert.theor:eq.Gamma}, $ \mathcal{W}[H] $ is the generating functional of cumulants, and $ \Psi = \{\Phi,\Phi'\} $ is the set of `classical' fields conjugated to $ \varphi $ defined as

\begin{align}
    \Psi(\xx,t) = \frac{\delta \mathcal{W}[H]}{\delta H(\xx,t)}.
\end{align}
The effective action serves as a generating functional of one-particle irreducible functions (e.g. proper vertexes), which are found by variational derivatives with respect to corresponding number of $ \Psi $ fields at $ \Psi = 0 $
\begin{align}
    \G^{\Psi...\Psi}(\{\xx_{i},t_{i}\}_{i=1}^{n}) = \frac{\delta^{n}\G[\Psi]}{\delta\Psi(\xx_{1},t_{1})...\delta\Psi(\xx_{n},t_{n})} \bigg|_{\Psi = 0}  . \label{S2:eq.GammaXY}
\end{align}
Furthermore, the effective action can be calculated from the response functional via well-known loop-expansion relation
\begin{align}
\G[\Psi] = - \s[\Psi] + \frac{1}{2} \text{Tr} \ln \frac{\delta \s[\Psi]}{\delta \Psi \delta \Psi} + \dots  \label{S2:eq.GS}.
\end{align}
The above formula can be proven for example using the saddle point approximation \cite{Vasilev2004,Amit2005,Zinn}. In Eq. \eqref{S2:eq.GS} the first term represents the tree (loopless) contribution, while the second term contains all one-loop contributions. The ellipsis denote higher-loop corrections which are irrelevant in the leading order. The one-loop corrections can be expanded in terms of the coupling constants, which gives
\begin{widetext}
	\begin{align}
	\frac{1}{2} \text{Tr} \ln \frac{\delta \s[\Psi]}{\delta \Psi \delta \Psi} =& \ \ \raisebox{-0.4cm}{\includegraphics[height=1cm]{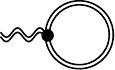}} \ + \ \raisebox{-0.4cm}{\includegraphics[height=1cm]{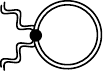}} \ + \ \raisebox{-0.4cm}{\includegraphics[height=1cm]{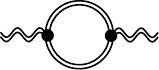}} \ + \ \raisebox{-0.4cm}{\includegraphics[height=1cm]{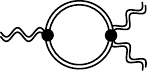}} \ + \  \raisebox{-0.4cm}{\includegraphics[height=1cm]{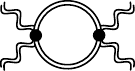}} \nonumber \\
	&\ + \ \raisebox{-0.55cm}{\includegraphics[height=1.3cm]{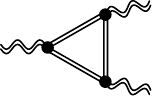}} \ + \ \raisebox{-0.55cm}{\includegraphics[height=1.3cm]{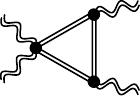}} \ + \  \raisebox{-0.55cm}{\includegraphics[height=1.3cm]{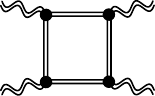}} + \dots \label{S2:eq.Gexpansion}
	\end{align}
\end{widetext}
where the ellipsis denote higher orders that are UV finite (see below). For the sake of simplicity, in \eqref{S2:eq.Gexpansion} we have only listed all possible topologies: the double solid lines represent propagators constructed from any of the $ \varphi = \{ \phi,\phi' \} $ fields, and double wiggly lines any of the $ \Psi = \{ \Phi,\Phi' \} $ lines, all of which are allowed by the structure of the present perturbation theory \eqref{S2:eq.p1}-\eqref{S2:eq.v2}.  For example, the third diagram in \eqref{S2:eq.Gexpansion} denotes a schematic representation of the following diagrams
\begin{align}
\raisebox{-0.4cm}{\includegraphics[height=1cm]{G_d3.pdf}} =&\ \frac{1}{4} \ \raisebox{-0.4cm}{\includegraphics[height=1cm]{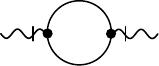}} +  \raisebox{-0.4cm}{\includegraphics[height=1cm]{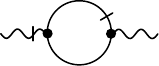}} \nonumber \\
&\ + \frac{1}{2} \ \raisebox{-0.4cm}{\includegraphics[height=1cm]{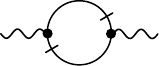}} \ , \label{S2:eq.G2explicit}
\end{align}

In the Fourier space, the loop contributions are constructed using the Feynman rules \eqref{S2:eq.p1}-\eqref{S2:eq.v2}, where every closed loop represents a single momentum and frequency integral (in $ d+d_{\omega} $ dimensional space). Note also, that all diagrams that contain closed loops of retarded propagators, such as the last diagram in \eqref{S2:eq.G2explicit}, vanish after the frequency integration (all poles in the propagators are present on a single side of the complex plane).

\section{Renormalized field theory} \label{S3:}

Continuum field theories are plagued with two types of divergences: infrared (IR) divergences for $ k \rightarrow 0 $, and ultraviolet (UV) divergences for $ k \rightarrow \infty $. Over the past decades, a systematic way of analysing properties of field theories on different scales based on UV renormalization have been developed, which represent the core idea of the field-theoretic RG \cite{Vasilev2004,Zinn,Amit2005}. As in the vast majority of the recent literature the details about the renormalization are absent, in the present work we will try to describe them in a rather pedagogical way.

\subsection{Canonical dimension in dynamical systems} \label{S3:A0}

In the RG approach to critical phenomena, we are ultimately interested in the critical scaling of the system. It is therefore important to understand the properties of the model under the change of the scale. Scaling (or engineering) dimensions of fields and parameters are defined by the condition that the response functional \eqref{S3:eq.RF} is invariant under two independent scale transformations: dilatation of the spatial coordinates $\xx\to a^{-1}\xx$, to which corresponds the transformation of the momentum coordinate $ \kk \rightarrow a \kk $, and dilatation of the time $t\to b^{-1}t$, to which corresponds the transformation of the frequency $ \omega \rightarrow b \omega $, where $ a,b $ are some dimensionless parameters. Therefore, for any quantity $ Q $ one can introduce  \emph{two} canonical dimensions: the momentum dimension $ d_{Q}^{k} $, and the frequency dimension $ d_{Q}^{\omega} $. For instance, field variables are scaled with respect to time by the rule $\phi(\xx,t)\to b^{d_{\phi}^{\omega}}\phi(\xx,b^{-1}t)$, and invariance of the time-derivative term in \eqref{S3:eq.RFa} gives rise to relation $d_{\phi}^{\omega}+d_{\phi'}^{\omega}=0$.

Canonical dimensions for the model \eqref{S3:eq.RF} are listed in Tab. \ref{tab:canon_CKPZ86}. As we are interested in the properties of the model within the simultaneous change of momentum and frequency scales, we define the total canonical dimension as \cite{Vasilev2004}
\begin{align}
d_{Q} = d_{Q}^{k} + d_{\omega} d_{Q}^{\omega}. \label{3A:eq.GHc}
\end{align}
The parameter $ d_{\omega} $ depends on the particular scale we are interested in, and its chosen in accordance with the underlying dispersion law $ \omega \sim k^{d_{\omega}} $ (in the standard literature $ d_{\omega} = z $ \cite{hohenberg1977theory,Tauber12}). In the most of the dynamical RG literature, $ d_{\omega} = 2 $, which allows us the study the IR properties of the model in the `diffusive' limit $ \omega \sim k^{2} \rightarrow 0 $. In the present work, we are mainly interested in the properties of the system in the limit $ \omega \sim k^{4} \rightarrow 0 $, hence we chose $ d_{\omega} = 4 $ (Similar scenario appears in Models B and H of critical dynamics \cite{hohenberg1977theory,Vasilev2004}). 

\begin{table}
	\def\arraystretch{1.3}
	\centering
	\begin{tabular}{c |c c c c c c c c }
		\hline \hline
		$Q$ & $ \phi $ & $ \phi' $ & $ D_{0},D $ & $ \tau_{0},\tau $ & $ \Lambda,\mu $ & $ \lambda_{0},\zeta_{0} $ & $ \eta_{0} $ & $ \lambda,\zeta,\eta $
		\\  \hline
		$d_Q^k$ & $ \tfrac{d-2}{2} $ & $ \tfrac{d+2}{2} $ & $ 4 $ & $ 2 $ & $ 1 $ & $ \tfrac{2-d}{2} $ & $ 2-d $ & $ 0 $
		\\ 
		$d^\omega_Q$ & $ 0 $ & $ 0 $ & $ -1 $ & $ 0 $ & $ 0 $ & $ 0 $ & $ 0 $ & $ 0 $
		\\ 
		$d_Q$ & $ \tfrac{d-2}{2} $ & $ \tfrac{d+2}{2} $ & $ 0 $ & $ 2 $ & $ 1 $ & $ \tfrac{2-d}{2} $ & $ 2-d $ & $ 0 $
		\\ \hline \hline    
	\end{tabular}
	\caption{Canonical dimensions of the bare fields and bare parameters for the response functional \eqref{S3:eq.RF}. Note that as we have rescaled all parameters with $ D_{0} $, none of them has non-zero frequency dimension.}
	\label{tab:canon_CKPZ86}
\end{table}

The canonical dimensions for $ d_{\omega} = 4 $ are listed in Tab. \ref{tab:canon_CKPZ86}. At this scale, all expansion parameters are dimensionless at $ d = 2 $, which implies that $ d_{c} = 2 $ is the upper critical dimension of the system. For $ d > 2 $, the non-linearities should be irrelevant and the `mean-field' exponents (i.e. canonical dimensions from Tab. \ref{tab:canon_CKPZ86}) should accurately describe the system. For $ d < 2 $, on the other hand, the corrections to scaling exponents coming from non-linearities must be taken into account. Moreover, the parameter $ \tau $ has a positive canonical dimension $ d_{\tau} = 2 $ at any $ d $, and so it will grow under the RG transformation unless $ \tau \approx 0 $ (it becomes a `mass'). Hence, the model is critical only in the massless limit $ \tau \rightarrow 0 $, where the perturbative RG analysis will be performed. Away from the critical point however, a crossover to a different scaling regime occurs: if $ \tau $ is not small, the leading IR dispersion law is $ \omega \sim k^{2} $ instead of $ \omega \sim k^{4} $. This type of asymptotic shows a trivial IR behavior, that will be later discussed in Sec. \ref{S6:D}.

It is important to mention the special r\^ole of the diffusion constant $ D_{0} $: Although it is dimensionless in the sense of the $ \omega \sim k^{4} $ scaling, it still has non-zero momentum and frequency canonical dimensions. Hence, although $ D_{0} $ does not grow under the scale transformation, i.e. it does not have to be small in the critical regime, it is not an expansion parameter (the same applies to $ \tau_0 $ in the $ \omega \sim k^{2} $ regime). We will further discuss the properties of this parameter in the IR limit in Sec. \ref{S4:A}. Note, that $ D_{0} $ is the only parameter with these special properties, as all other parameters were rescaled with it. 

\subsection{Degree of divergence} \label{S3:A}

The starting point for the renormalization analysis is the UV index of divergence. In general, any one-particle irreducible Feynman diagram contributing to \eqref{S2:eq.GammaXY} with $ N $ number of external fields  can be written as  $ (-g_{0})^{n} I_{nN} $, where $ n $ is the order of the perturbation theory, $ g_{0} $ is a product of coupling constants such that $ [g_{0}] = [\mu^{2-d}] $, and the factor $ I_{nN} $  contains the remaining frequency and momentum integrals.  The degree of divergence is then the dimension of $ I_{nN} $, which can be obtained from the topological properties of the perturbation theory (see Appendix \ref{APP:DoD})
\begin{align}
	d[I_{nN}] =&\ d + 4 - \tfrac{1}{2} (d+2) N_{\phi'} - \tfrac{1}{2} (d-2) N_{\phi} \nonumber \\
	&\ - \tfrac{1}{2} (2-d) V_{\phi'\phi\phi} - (2-d) V_{\phi'\phi\phi\phi}. \label{S3:eq.dG}
\end{align}
In the above, $ N_{\phi'}, N_{\phi} $ are the number of external fields $ \phi',\phi $, and $ V_{\phi'\phi\phi},V_{\phi'\phi\phi\phi} $ are the number of three and four point vertexes. The expression \eqref{S3:eq.dG} can be understood by the means of simple dimensional analysis: the term $ d + 4 $ represent dimension of (suppressed) momentum and frequency conserving delta functions in the limit $ \omega \sim k^{4} $, the next two terms come from the variational derivatives \eqref{S2:eq.GammaXY}, and the remaining terms are just the dimensions of subtracted coupling constants. 

In general, any diagram with $ d[I_{nN}] \geq 0 $ might show UV divergence. From \eqref{S3:eq.dG} it is evident that for $ d > 2 $, the degree of divergence increases with the order of perturbation theory, and decreases for $ d < 2 $. Moreover, at $ d = 2 $ the UV divergence is independent of the order of the perturbation theory, and is determined solely from the number of external fields. The systematic UV renormalization process can be carried out only when the number of divergent structures is finite, i.e. for renormalizable ($ d = 2 $), and superrenormalizable theories ($ d < 2 $) \cite{Zinn,Vasilev2004}. For $ d > 2 $ every vertex function contains divergent contributions which makes the theory nonrenormalisable. 

It is instructive to describe the features of UV divergences on a simple one-loop integral. After the frequency integral has been carried out, the typical momentum integral encountered in the current perturbation theory has the following form 
\begin{widetext}
\begin{align}
\int_{|\kk|<\Lambda}  \ \frac{\dd \kk}{k^{2} + \tau}
=
\begin{cases} \pi\,
\ln \left( {\displaystyle \Lambda^{2}\over\displaystyle\tau}\right) + \mathcal{O}\left( {\displaystyle\tau\over\displaystyle\Lambda^{2}}\right), & 
\ d = 2, \\
\frac{\displaystyle 2\pi^{d\over 2}}{\displaystyle (2-d)\Gamma(\tfrac{d}{2})}\,\tau^{\tfrac{d-2}{2}} \Bigg\{\Gamma(\tfrac{d}{2})\Gamma(2-\tfrac{d}{2}) - \left({\displaystyle\Lambda^2\over\displaystyle\tau}\right)^{\tfrac{d-2}{2}}\left[1+ \mathcal{O}\left( {\displaystyle\tau\over\displaystyle\Lambda^{2}}\right)\right] \Bigg\}, &
\ d \ne 2.
\end{cases} \label{S3:eq.int}
\end{align} 
\end{widetext}
where $ \Lambda $ is the UV cut-off and $ \tau $ serves as an IR cut-off.
From (\ref{S3:eq.int})
it is clear that the integral is UV divergent for $ d > 2 $ when the `continuum' limit is taken $ \Lambda \rightarrow \infty $, and IR divergent for $ d < 2 $ as the system approaches the critical point $ \tau \rightarrow 0 $. As we are interested mainly in the IR properties (in general $ \tau \rightarrow 0 $ and $ \omega \sim k^{d_{\omega}} \rightarrow 0 $), no sensible results can be derived for $ d \leq 2 $ within \emph{any} finite order of the standard perturbation theory. Although there are currently \emph{no} available methods which would allow us to eliminate IR divergences at the critical point, the UV divergences can be eliminated via a systematic renormalization procedure \cite{Vasilev2004,Zinn,Amit2005}.
Connection between the UV divergence with the critical limit $\tau\to 0$ is seen in the dependence on the
ratio $\Lambda^2/\tau$ only of the function in the braces on the right side of (\ref{S3:eq.int}). The UV divergence is suppressed by the prefactor $\tau^{\tfrac{d-2}{2}}$ above two dimensions, but this does not happen at the upper critical dimension $d_{c}\equiv2$. At $ d = d_{c} $ the limit $ \Lambda \rightarrow \infty $ formally corresponds to the critical point limit $\tau\to 0$. This relation between UV and IR then allows us to study the IR limit by systematically analysing UV properties of the field-theoretic model~\footnote{In field-theoretic approach with local interactions, higher order gradients in the free Langevin equation spoil the UV-IR connection. Hence, they must be dropped.}.

The above example describes a very important idea behind the field-theoretic RG: While the IR universal quantities are independent of the microscopic properties of the system (such as the UV cut-off $ \Lambda $), all important information about critical scaling actually comes from UV renormalization (as in Wilson's RG). As will be shown later, at $ d = d_{c} $, the UV renormalization of \eqref{S3:eq.RFa} can be done by absorbing all divergences into the fields and finite number of parameters of the model. Once this is done, the model becomes scale dependent and the IR asymptotic properties (such as critical exponents) can be derived as a series expansion in terms of $ \varepsilon = d_{c} - d \ll 1 $. One must keep in mind however, and it will be also discussed below, that the whole perturbative RG analysis is justified only close to the upper critical dimension.

The vertex functions which might require renormalization are those, in which contributions from Feynman diagrams have non-negative UV exponent at the upper critical dimension
\begin{align}
\delta_{\G} \equiv d[I_{nN}] |_{d = 2} = 6 - 2 N_{\phi'}.
\end{align}
Note, that the above expression is completely independent of the number of $ \phi $ fields. This implies, that in principle there could be an infinite number of vertex functions (containing any powers of $ \phi $ fields), which would require renormalization. In other words, the theory is non-renormalizable. Similar situation appears in other field theories as well, but systematic perturbative renormalization in the case of stochastic processes was carried out only in few special cases \cite{Antonov1995,Antonov2017}.

In the present model however, other aspects of the perturbation theory must be also taken into account. One can see from the response functional \eqref{S3:eq.RF}, that the fields $ \phi' $ and $ \phi $ are always coupled with at least one $ \partial $ operator. This means that in the process of calculating Feynman diagrams the corresponding external momentum can be always taken out of the integral. Hence, the real degree of divergence $\delta_{\G}'$ is reduced by the total number of these fields $ \delta_{\G}' = \delta_{\G} - N_{\phi'} - N_{\phi} $, which gives
\begin{align}
\delta_{\G}' = 6 - 3 N_{\phi'} - N_{\phi}. \label{S3:eq.Gs}
\end{align}
Moreover, due to the retardation property of the propagators \eqref{S2:eq.p1}-\eqref{S2:eq.p2}, all vertex functions without at least one response field $ \phi' $ vanish (See Sec. \ref{S2:C} and \cite{Vasilev2004,Tauber12}). One can see from \eqref{S3:eq.Gs}, that there is now only a finite number of possible vertex functions which may require renormalization. These are the following
\begin{align}
\phi', \quad \phi'\phi' , \quad \phi'\phi, \quad \phi' \phi^{2}, \quad  \phi'\phi^{3}. \label{S3:eq.DV}
\end{align}
The same tremendous simplification of the renormalization process actually appears in the usual KPZ equation as well \cite{KPZ86}, where the field equations are obtained by the series expansion in terms of $ \partial h $, rather than the height field $ h $. Would the field $ h $ be decoupled from $ \partial $, the theory will be non-renormalizable \cite{Antonov1995}.

From the divergent vertex functions \eqref{S3:eq.DV} one can construct the following counter-terms consistent with the response functional \eqref{S3:eq.RF}
\begin{align}
&\partial^{2}\phi', \ \ \partial^{4}\phi' , \ \ (\partial\phi')^{2}, \ \ \phi'\partial_{t} \phi, \ \ \phi'\partial^{4} \phi, \ \ \phi' \tau \partial^{2} \phi, \\
&(\partial\phi)^{2}\partial^{2}\phi', \ \ ( \partial_{i} \phi' ) (\partial_{i} \phi) (\partial^{2} \phi), \ \ ( \partial_{i} \phi' ) (\partial_{i} \phi) (\partial\phi)^{2}.
\end{align}
An important remark is in order. The fact that one can construct counter-terms with a non-negative degree of divergence \eqref{S3:eq.Gs} does not yet mean that they will be unavoidably generated in the perturbation theory. For example, all contributions coming from the counter-terms $ \propto \phi' $ (which renormalize the mean value of the random noise $ \langle f \rangle $ \cite{Antonov2015}) have the following form
\begin{align}
\raisebox{-0.4cm}{\includegraphics[height=1cm]{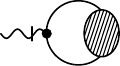}} \ , \quad \raisebox{-0.4cm}{\includegraphics[height=1cm]{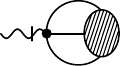}} \ ,
\end{align}
where shaded bubbles denote all possible sub-diagrams.
As all external vertexes are proportional to external momentum $ p $, these diagrams vanish due due to the conservation of the momentum. Similar ideas will be used in the next section in order to show that the models with cubic and quartic interactions are decoupled.

\subsection{Decoupling of the $ \phi'\phi^{2} $ and $ \phi'\phi^{3} $ models} \label{S3:B}

We have shown in the previous section that the model is renormalizable with both cubic and quartic interactions. The remaining question is: are the corrections to quartic/cubic interaction generated solely from the cubic/quartic interaction? If this is not the case, the models are decoupled and one can study each of them separately.

\emph{Decoupling of $ \phi'\phi^{3} $.} First, we notice that in the absence of cubic interactions $ \lambda = \zeta = 0 $ the model \eqref{S3:eq.RF} (and \eqref{S1:eq.LE}) posses the `Ising' symmetry
\begin{align}
\phi \rightarrow - \phi, \quad \phi' \rightarrow -\phi'.
\end{align}
This means, that divergent contributions to $ \phi'\phi\phi $ interaction cannot arise from four-point interaction only, in all orders of the perturbation theory.

\emph{Decoupling of $ \phi'\phi^{2} $.} What remains is to show that no divergent corrections to $ \phi'\phi^{3} $ interaction can arise from considering the $ \phi'\phi^{2} $ interaction only. This can be done by analysing the properties of the vertex factors. As we have mentioned in the previous section, if $ \partial $ operators are attached to fields in the interaction terms, the corresponding number of external momenta can be taken out of the diagram, which reduces its degree of divergence. Moreover, the present model has additional special feature. Consider a correction to one of the vertex functions, in which an external $ \phi' $ field is connected to a three-point vertex \eqref{S2:eq.v1}. Denoting the external and internal momenta as $ \pp $ and $ \kk $, the corresponding vertex factor is
\begin{align}
\raisebox{-0.53cm}{\includegraphics[height=1cm]{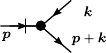}} &= D_{0} [ \lambda_{0} k^{2} p^{2} + \tfrac{1}{2} \zeta_{0} (p^{2}k^{2} - 2(\pp.\kk)^{2}) ] \nonumber \\
&+  \mathcal{O}(p^{3}). 
\end{align} 
As we can see, every time an external $ \phi' $ field is attached to a three point vertex, the degree of divergence is actually reduced by $ 2 $! This implies, that any diagram contributing to the noise $ \phi'^{2} $, and to four-point vertex $ \phi'\phi^{3} $ constructed solely from three-point vertexes will have the following form
\begin{align}
\raisebox{-0.3cm}{\includegraphics[height=0.8cm]{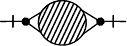}} = \mathcal{O}(p^{4}) \ , \quad \raisebox{-0.5cm}{\includegraphics[height=1.2cm]{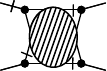}} = \mathcal{O}(p^{5}) \ , \label{S3:eq.diagsO45}
\end{align}
where shaded bubbles denote all possible sub-diagrams. As the marginally divergent terms for the above diagrams are of order $ p^{2} $ and $ p^{4} $ respectively, any of such diagrams will be finite. Similar situation simplifying the renormalization process occurs in turbulent advection of active scalar field \cite{Nandy_1998,antonov2019}

The above results also imply, that both noise and mass renormalization are absent in the model without the quartic interaction to all orders of perturbation theory, and therefore additional relation between scaling exponents should hold (as it was in the original CKPZ model \cite{Janssen1997}). Inclusion of the quartic interaction will however generate logarithmic corrections to noise term (starting from two loops) and both quadratic and logarithmic corrections to mass term.

\subsection{Renormalization} \label{S3:D}

We have mentioned in Sec. \ref{S3:A} that both UV and IR cut-offs must be included in order to ensure the finiteness of the theory. The UV divergences can be then eliminated by performing systematic renormalization procedure, and the connection between UV and IR limits allows us to study IR asymptotic. For practical calculations however, it is inconvenient to work with a sharp form of cut-off. This is because the cut-offs regularize \emph{all} propagators involved in the perturbative expansion, and \emph{not} the loop integrals. In the present model, the regularized propagators have the following form \cite{Vasilev2004,Zinn}
\begin{align}
\langle \phi \phi' \rangle_{0}|_{\text{reg}} = \frac{\theta(\Lambda - k) - \theta(m-k)}{-i\omega + D_{0}k^{2}(k^{2}+\tau_{0})}, \label{S3:eq.P1} \\
\langle \phi \phi \rangle_{0}|_{\text{reg}} = \frac{D_{0}k^{2} [\theta(\Lambda - k) - \theta(m-k) ]}{\omega^{2} + D_{0}k^{4}(k^{2}+\tau_{0})^{2}}, \label{S3:eq.P2}
\end{align}
where $ m $ is the IR cut-off parameter, and $ \theta(...) $ denotes either Heaviside step function, or a sufficiently fast decaying function around $ k \sim \Lambda,m $ \cite{Zinn}. The origin of the regularization \eqref{S3:eq.P1}-\eqref{S3:eq.P2} can be tracked back to the regularized Fourier-space representation of the fields
\begin{align}
\phi(\xx,t) = \int_{m<|\kk|<\Lambda} \dd \kk \ \phi(\kk,t).
\end{align} 
It should be also noted that in the wast majority of the literature IR finitness is provided by the deviation from the criticality - in our case the integrals are IR finite due to $ \tau > 0 $, so that one may safely take the limit $ m \rightarrow 0 $. This choice is, however, only the matter of convenience: The exact form of IR cut-off does not change the universal quantities, provided the fact that it regulates all IR divergences systematically to all orders of perturbation theory (and that its small in the IR limit). There are also other ways of ensuring IR finiteness of the integrals (non-zero external momentum/frequency \cite{Tauber12}, or cut-off in the noise correlator \cite{Vasilev2004}), but the most natural IR regulator for the model \eqref{S3:eq.RF} is the deviation from the criticality $ \tau $.  

The above apparently minor change of course does not normally spoil the leading UV behavior of the one-loop diagrams, and hence the leading critical scaling when calculated in the context of $ \varepsilon $-expansion. However, in the dynamical RG literature several authors  tend to work with the `fixed $ d $ renormalization group', where the entire calculation is carried out in a fixed dimension $ d $ \cite{KPZ86,FreyTauber94,Haselwandter2008}. In such case, UV finite terms of the one-loop diagrams can be included, but only if calculated in a consistent way with propagators \eqref{S3:eq.P1} and \eqref{S3:eq.P2}. Note, that similar ideas apply to Wilson's RG, where the momentum integrals are in general \emph{not} restricted to a simple momentum shell $ \Lambda/b < |\kk| < \Lambda $, but the precise shape of the `shell' is determined by propagators \cite{Wilson1974}. We will further discuss Wilson's RG in Sec. \ref{S4.E} when comparing our results with previous works.

In order to avoid the above-mentioned difficulties coming from a cut-off regularization, it is convenient to chose a different approach, in which calculations are carried out without a UV cutoff. For instance, little reflection shows that at fixed cutoffs the integral on the left side of (\ref{S3:eq.int}) is a regular function of $d$ on the right half of the complex $d$ plane: $d>0$. In dimensional renormalization this function is analytically continued to the whole $d$ plane with singularities on the left half plane~\cite{THOOFT1972189,*tHooft1973}. Expansion of the integral (\ref{S3:eq.int}) in the parameter $\tau$, however, gives rise to terms with singularities on the right half plane $d$, as seen from the expressions on the right side of (\ref{S3:eq.int}). In particular, when $d<2$ the limit $\Lambda\to \infty$ may be taken and the resulting expression contains now a pole at $d=2$ instead of the logarithmic divergence at $d=2$. This is wherefrom the tradeoff comes between UV singularities in terms of the cutoff $\Lambda$ and the singularity in space dimension $1/(2-d)$.

The main idea behind dimensional renormalization 
can be intuitively understood in the following way \cite{Vasilev2004,Zinn,Amit2005}. Consider the model \eqref{S3:eq.RF} in $ d = 2 - \varepsilon $ dimension with the UV cutoff $ \Lambda $, with $ e_{0} $ being the set of all bare parameters.  First, the quadratic divergences responsible for the shift of the critical point are eliminated by an \emph{additive renormalization} of the mass parameter $ \tau_{0} $. In the one loop approximation this can be done by the following substitution
\begin{align}
\tau_{0} + c(\varepsilon) g_{0} \Lambda^{2-\varepsilon} \rightarrow \tau_{0}'(\Lambda),
\end{align}
where $ c(\varepsilon) $ is a regular function at $ \varepsilon = 0 $. This $\Lambda$-dependent renormalization does not affect the scaling behavior of the deviation $\Delta \tau$ of the mass parameter from its critical value. Upon this $\Lambda$ renormalization all remaining integrals are finite in the limit $ \Lambda \rightarrow \infty $, which then can be performed as long as $ \varepsilon > 0 $. Passing to this limit has no effect on the large-scale behavior of the model, which is independent of the microscopic details, but renders wave-vector integrals easier to handle, especially in high orders of perturbation theory \cite{Vasilev2004}.
The remaining $ 1/\varepsilon $
poles
are eliminated with the \emph{multiplicative renormalization} of all parameters $ e_{0}' = Z_{e} e $, and fields $ \varphi \rightarrow Z_{\varphi} \varphi $. The full renormalization procedure can be schematically written  as \cite{Vasilev2004}
\begin{align}
\{ e_{0}, \varphi \} \xrightarrow{\Lambda-\text{ren}} \{ e_{0}' , \varphi \} \xrightarrow{\varepsilon-\text{ren}} \{ e, Z_{\varphi} \varphi \}.
\end{align} 
It should be noted that, in general, the parameters $e_0'$ depend on $\Lambda$ and are different form the original bare parameters $e_0$. To simplify notation, in what follows we shall use the notation $e_0$ for the bare parameters obtained as the result of the $\Lambda$ renormalization, because the actual values of bare parameters are irrelevant in calculation of anomalous scaling exponents.

\subsection{Renormalized response functional}

As the length scale $ \Lambda $ disappears from the model in the limit $\Lambda\to\infty$, a new dimensional parameter is introduced in order to make the coupling constants dimensionless \cite{Vasilev2004,Zinn,Amit2005}: an auxiliary mass scale $ \mu, $ where $ [\mu] = [\Lambda] $. The renormalized field-theoretic model is then obtained by renormalizing the parameters in the following way
\begin{align}
D_{0} &= D Z_{D}, \quad \tau_{0} = Z_{\tau} \tau + \tau_{c} \quad \lambda_{0} = \lambda \mu^{\varepsilon/2} Z_{\lambda}, \label{S4:eq.renPar1} \\
\zeta_{0} &= \mu^{\varepsilon/2} \zeta Z_{\zeta}, \quad  \eta_{0} = \mu^{\varepsilon} \eta Z_{\eta}, \label{S4:eq.renPar2}
\end{align}
together with the field renormalization $ \varphi \rightarrow Z_{\varphi} \varphi $. The term $ \tau_{c} $ denotes the additive renormalization of the mass parameter $ \tau_{0} $, which comes from the $\Lambda$ renormalization and leads to the substitution $\tau_0\to \tau$ in propagators of the renormalized perturbation theory.
The renormalized response functional \eqref{S3:eq.RF} then becomes
\begin{align}
\s[\varphi] =&\  \tfrac{1}{2} Z_{1} D \phi'\partial^{2}
\phi' + \phi' \{ \partial_{t} + D (Z_{2} \partial^{4} - Z_{6} \tau \partial^{2} )  \} \phi \nonumber \\
&\ + \tfrac{1}{2} D Z_{3} \mu^{\varepsilon/2} \lambda (\partial^{2}\phi') (\partial \phi)^{2} \nonumber \\
&\ + \tfrac{1}{2} D Z_{4}  \mu^{\varepsilon/2} \zeta ( \partial_{i} \phi' ) (\partial_{i} \phi) (\partial^{2} \phi) \nonumber \\
&\ + \tfrac{1}{3!} D Z_{5} \mu^{\varepsilon} \eta (\partial_{i} \phi')(\partial_{i}\phi)(\partial\phi)^{2} , \label{3:eq.RSF}
\end{align}
where 
\begin{align}
Z_{\phi} &= ( Z_{1}^{-1} Z_{2} )^{\frac{1}{2}} , \quad Z_{\phi'} = ( Z_{1} Z_{2}^{-1} )^{\frac{1}{2}} , \quad Z_{D} = Z_{2}, \label{S4:eq.Zc} \\
Z_{\lambda} &= ( Z_{1} Z_{2}^{-3}  )^{\frac{1}{2}} Z_{3}, \quad Z_{\zeta} = ( Z_{1} Z_{2}^{-3}  )^{\frac{1}{2}} Z_{4}, \\
Z_{\eta} &= Z_{1} Z_{2}^{-2} Z_{5}. \quad Z_{\tau} = Z_{2}^{-1}Z_{6}. \label{S4:eq.Zd}
\end{align}
The divergent corrections to the vertex function have the following form
\begin{widetext}
\begin{align}
\G_{\phi'\phi'} =&\ D p^{2} Z_{1} + \dots , \label{S4:eq.corr2Point} \\
\G_{\phi'\phi} =&\ i\Omega - Z_{2} D p^{4} + Z_{6} D \tau p^{2} + \ \raisebox{-0.34cm}{\includegraphics[height=0.9cm]{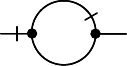}} + \frac{1}{2} \ \raisebox{-0.17cm}{\includegraphics[height=0.65cm]{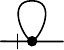}} + \dots , \\
\G_{\phi_{\pp+\rr}'\phi_{-\pp}\phi_{-\rr}} =&\  - D \mu^{\varepsilon/2} \lambda Z_{3} \pp.\rr(\pp+\rr)^{2} + \tfrac{1}{2} D \mu^{\varepsilon/2} \zeta Z_{4} [2 p^{2}r^{2} + (r^{2} + p^{2})\pp.\rr ] + \nonumber \\
&\ + \ \raisebox{-0.48cm}{\includegraphics[height=1.2cm]{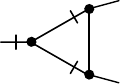}} + 2 \ \raisebox{-0.48cm}{\includegraphics[height=1.2cm]{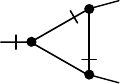}} + \ \raisebox{-0.34cm}{\includegraphics[height=0.9cm]{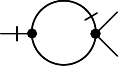}} + 2 \ \raisebox{-0.34cm}{\includegraphics[height=0.9cm]{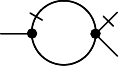}} + \dots , \label{S4:eq.corr3Point} \\
\G_{\phi_{\pp+\qq+\rr}'\phi_{-\pp}\phi_{-\qq}\phi_{-\rr}} =&\ \tfrac{1}{3} D \mu^{\varepsilon} \eta Z_{5} \big[ p^2 (\qq.\rr) + (\pp.\rr) \left(q^2 + 2 (\qq.\rr)\right) + (\pp.\qq) \left(2 (\pp.\rr)+2 (\qq.\rr) + r^2\right) \big] \nonumber \\
& + 3 \ \raisebox{-0.34cm}{\includegraphics[height=0.9cm]{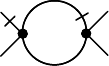}} + 3 \ \raisebox{-0.48cm}{\includegraphics[height=1.2cm]{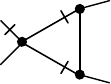}} + 6 \ \raisebox{-0.48cm}{\includegraphics[height=1.2cm]{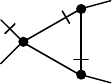}} + \dots , \label{S4:eq.corr4Point}
\end{align}
\end{widetext}
where we have listed only divergent Feynman diagrams, and their calculation is described in Appendix \ref{APP:FD}.

Once the diagrams have been calculated the renormalization constants are extracted using a particular \emph{subtraction scheme} \cite{collins_1984,Vasilev2004,Zinn,Amit2005}. The two most commonly used are the \emph{normalization point} (NP) and \emph{minimal subtraction} (MS) schemes, which we will both discuss now. In the NP scheme, the renormalization constants are determined from the normalization conditions in which all potential IR cut-off parameters are fixed at a specific normalization point. In the present model, this can be achieved by employing the following normalisation conditions
\begin{align}
\partial_{p^{2}}\G_{\phi'\phi'}|_{\text{n.p.}} =&\ D, \label{S3:eq.NC1} \\
\partial_{p^{4}}\G_{\phi'\phi}|_{\text{n.p.}} =&\ -D, \\
\partial_{p^{2}} \partial_{\tau}\G_{\phi'\phi}|_{\text{n.p.}} =&\ D, \\
\partial_{p_{i}}\partial_{p_{i}}\partial_{r_{j}}\partial_{r_{j}} \G_{\phi'\phi\phi}|_{\text{n.p.}} =&\ 4d D \mu^{\varepsilon/2} (d\zeta-2\lambda) \\
\partial_{p_{i}}\partial_{p_{i}}\partial_{q_{j}}\partial_{r_{j}} \G_{\phi'\phi\phi\phi}|_{\text{n.p.}} =&\ \tfrac{2}{3} d(d+2) D \mu^{\varepsilon} \eta, \label{S3:eq.NC2}
\end{align}
where $ \text{n.p.} = \{ \tau = \mu^{2}, \ \Omega_{i} = 0, \ p_{i} = 0, \ \forall i \} $ stands for the normalization point. The normalisation conditions \eqref{S3:eq.NC1}-\eqref{S3:eq.NC2} have been chosen for convenience, but the physical predictions are independent of their precise form (universal properties are `renorm-invariant').  The one-loop renormalization constants then attain the following form
\begin{align}
Z_{1} &= 1, \label{S4:eq.Za} \\
Z_{2} &= 1 - \bigg( \frac{\lambda  (\lambda - 2 \zeta )}{4 (2-d)} \nonumber \\
&\ \hspace{1.2cm} + \frac{(\zeta + 2 \lambda ) ((7 d+8) \zeta + 2(d+2) \lambda )}{64 (d+2)} \bigg), \\
Z_{3} &= 1 + \frac{\eta}{2} \left( \frac{1}{2-d} -\frac{(d+3) \zeta +2(d+1) \lambda }{12 (d+2) \lambda } \right), \\
Z_{4} &= 1 - \frac{\eta}{12\zeta} \left( \frac{2\lambda -7 \zeta }{2-d} +\frac{(7 d+12) \zeta +2(d+4) \lambda }{4 (d+2)} \right), \\
Z_{5} &= 1 + \frac{\eta}{4} \left( \frac{3}{2-d} - \frac{d-1}{3(2+d)} \right), \\
Z_{6} &= 1 - \frac{\eta}{12} \left( \frac{4}{2-d} - 1 \right). \label{S4:eq.Zb}
\end{align}
Note the appearance of the UV finite terms at $ d = 2 $ in the renormalization constants. In the NP scheme, there is \emph{no} reason for discarding them, but neither it is forbidden to do so. The point is, that if one is interested in calculating the scaling exponents in the framework of the $ \varepsilon $-expansion, these finite terms will not create any contributions. This is the main idea behind the MS scheme, where the renormalization constants only minimally subtract terms that are divergent in the limit $ d \rightarrow d_{c} $ \footnote{In MS scheme, it is common to rescale coupling constants with some regular function of $ d $. This small change does not influence the universal properties of the system, and it is sometimes referred to as $\overline{\text{MS}}$ scheme.}. Note, that the MS scheme does \emph{not} require specification of any normalization conditions such as \eqref{S3:eq.NC1}-\eqref{S3:eq.NC2}. Instead, the renormalization constants are obtained from performing series expansion of all $ d $-factors in terms of $ \varepsilon $, and from the requirement that the vertex functions \eqref{S4:eq.corr2Point}-\eqref{S4:eq.corr4Point} are UV finite for $ \varepsilon \rightarrow 0 $, at any momenta, frequency etc. \cite{collins_1984,Vasilev2004,Zinn,Amit2005}.

Although the MS scheme is the most convenient for practical calculations, such as multi-loop calculations \cite{kleinert2001critical}, it has its limitations \cite{FreyTauber94,Honkonen2017}. In the present model, we are also interested in the $ d = 1 $ behavior. In the context of $ \varepsilon $-expansion, this can be done calculating all quantities as a series of $ \varepsilon $, and by performing analytical continuation to $ \varepsilon \rightarrow 1 $ at the end of the day. The advantage of the NP scheme on the other hand is, that we can evaluate all quantities in a fixed dimension, and therefore study the properties of $ d = 1 $ system directly. This has a particular advantage in the present model, as we know that in $ d = 1 $ the two cubic interactions merge into one (see Sec. \ref{S2}). Additional relation between normalisation constants then appears, which will be further discussed in Sec. \ref{S4:d1}. If we would use the MS scheme, this information will be lost. The similar ideas were employed in the two-loop calculation of the standard KPZ equation \cite{FreyTauber94}, where the NP scheme was used in order to preserve the time-reversal symmetry (i.e. the fluctuation-dissipation theorem) in $ d = 1 $ \footnote{As it was already pointed out by Wiese \cite{Wiese1997} one  should not misinterpret the argumentation in \cite{FreyTauber94}; the so-called `geometric factors' must \emph{not} be distinguished from the dimensional factors appearing in the expansion parameter $ \varepsilon = 2 - d $ of the space, but the NP scheme allows one to keep the finite terms of the diagrams in renormalization constants.}.

It is very important, however, to keep in mind that the $ \varepsilon $-expansion is the \emph{only} reliable framework for performing perturbative RG analysis, because: $ i) $ the multiplicative renormalization, and the connection between IR and UV holds only at $ d = d_{c} $, $ ii) $ some composite structures might become relevant if $ \varepsilon $ is not small \cite{Vasilev2004,Caballero2020}, and $ iii) $ there is no other small parameter besides $ \varepsilon $ that can be used for calculation of scaling exponents at a fixed point~\footnote{The fixed $d$ procedure can be justified for example if a different expansion parameter is used as well that ensures smallness of coupling constants at the fixed point. This is the case for example in the double-expansion scheme of fully developed incompressible turbulence \cite{Adzhemyan2010}}. Moreover, if the $ \varepsilon $ expansion fails to produce a perturbative fixed point for $ 0 < \varepsilon \ll 1 $, but the fixed $ d $ calculation shows a stable IR fixed point at $ d = 1 $, one \emph{cannot} expect that the latter will describe a physical scenario. An exemplary case of this situation is the standard KPZ equation: while the fixed $ d = 1 $ one-loop calculation apparently produces a perturbative fixed point with exact scaling exponents \cite{KPZ86}, the resummed perturbative expansion shows that \emph{no} fixed point exists for any $ d < 2 $ \cite{Wiese_1998} (the dynamics in $ d = 1 $ is driven by a non-perturbative fixed point \cite{Canet2010,Canet2011}). Further discussion about fixed $ d $ RG can be found for example in \cite{Vasilev2004}, Chapter 1.41.

\section{Critical Scaling} \label{S4}

We are mainly interested in deriving scaling properties of the model \eqref{S3:eq.RFa} in the critical region $ \tau \rightarrow 0 $, and $ \omega \sim k^{4} \rightarrow 0 $. As classical perturbative expansion is divergent in this limit, one must use RG technique in order to resum the entire series. We will now derive and solve the RG equation for a general two-point correlation function, and then further apply this method to $ d = 2 - \varepsilon $ and $ d = 1 $ cases separately.

\subsection{RG equation} \label{S4:A}

As we have already renormalized the model, we can study the effective properties of the correlation functions depending on the scale of interest. In approaches based on the momentum-shell or exact renormalization group, where the
field-theoretic models are renormalized 
using infinitesimal changes of
the UV cut-off $ \Lambda $, the RG equation is derived by looking at the change of correlation functions w.r.t. $ \Lambda $. The universal properties are then obtained by finding the IR fixed point in the limit $ k/\Lambda, i\omega/D\Lambda^4, \tau/\Lambda^2 \rightarrow 0 $, while keeping $ \Lambda $ fixed.  However, in dimensional renormalization, where the limit $ \Lambda \rightarrow \infty $ was taken before the renormalization process, this approach is not possible anymore. Instead, the RG equations is derived by varying the introduced parameter $ \mu $, and the IR fixed point is found in the limit $ k/\mu, i\omega/D\mu^4, \tau/\mu^2 \rightarrow 0 $.

By knowing all renormalization constants, one can write the relation between bare and renormalized correlation functions as

\begin{align}
G^{(N_{\phi},N_{\phi'})}(\{ \kk_{i},\omega_{i} \},D_{0},\tau_{0},g_{0}) =&\ \nonumber \\
&\ \hspace{-3.75cm} = Z_{\phi}^{N_{\phi}} Z_{\phi'}^{N_{\phi'}} G^{(N_{\phi},N_{\phi'})}(\{ \kk_{i},\omega_{i}\},D,\tau,g,\mu), \label{S4:eq.GG}
\end{align}
where $ g = \{ \lambda,\zeta,\eta \} $ again denotes set of all coupling constants and $ g_{0} $ are their bare counter-parts. We are interested in how does the model change if we vary the mass scale $ \mu $. The bare theory is independent of $ \mu $.
Let us denote the logarithmic derivative with respect to any quantity $ x $ as $ \D_{x} = x\partial_{x} $, and the corresponding operation with keeping all bare parameters fixed as $ \D_{\mu}|_{0} = \mu \partial_{\mu} |_{g_{0}} $. Applying the latter operator to \eqref{S4:eq.GG} we derive the differential RG equation
\begin{align}
\big[ \D_{\mu} + \beta_{g} & \partial_{g} - \gamma_{D} \D_{D} - \gamma_{\tau} \D_{\tau} + N_{\phi'}\gamma_{\phi'} + N_{\phi}\gamma_{\phi} \big] \nonumber \\
&\ \times G^{(N_{\phi},N_{\phi'})}(\{\kk_{i},\omega_{i}\},D,\tau,g,\mu) = 0, \label{S4:eq.RGEq}
\end{align}
where we have defined the RG flow equations $ \beta_{g} $ and anomalous dimensions $ \gamma_{i} $ as
\begin{align}
\beta_{g} = - g (d_{g} + \gamma_{g}), \quad \gamma_{i} = \D_{\mu}|_{0} \ln Z_{i}. \label{S4:eq.BetGgam}
\end{align}
The explicit form of these functions will be discussed in the following subsections. Eq. \eqref{S4:eq.RGEq} describes the effective properties of the system depending on the scale of observation. Note, that in general one can study effective scaling `both ways', i.e. to start with the model on small scales and study larger scales, or vice versa. In other words, Eq. \eqref{S4:eq.RGEq} shows the group character of the RG transformation \cite{Bogolyubov1955} (as opposed to Wilson's RG, which is only a semi-group).

We will now solve the RG equation \eqref{S4:eq.RGEq} for the case of two point correlation function, and the general solution can be found in \cite{Vasilev2004,Tauber12}. From the dimensional considerations, the momentum-frequency representation of the renormalized two-point correlator is
\begin{align}
G^{(2,0)}(k,\omega,D,\tau,g,\mu) = k^{2d_{\phi}^{k}-d-d_{\omega}} D^{-1} C(s,u,z,g), \label{S4:eq.C}
\end{align}
where we have introduced the following dimensionless parameters
\begin{align}
s = k/\mu, \quad u = i\omega/D\mu^{4}, \quad z = \tau/\mu^{2}.
\end{align}
Substitution of \eqref{S4:eq.C} into the RG equation \eqref{S4:eq.RGEq} leads to an equation for $ C $
\begin{align}
\big[ -\D_{s} + \beta_{g} \partial_{g} - (&2-\gamma_{D}) \D_{u} - (2+\gamma_{\tau}) \D_{z} \nonumber \\
& \hspace{-0.2cm} + 2 \gamma_{\phi} + \gamma_{D} \big] \times C(s,u,z,g) = 0.
\end{align}
This equation can be solved in the standard fasion using the method of characteristics \cite{Vasilev2004}. In the present case, we have five invariant variables
\begin{align}
\D_{s} \bar{g} &= \beta_{g}(\bar{g}), \hspace{1.8cm} \bar{g}(s=1) = g_{I}, \label{S4:eq.IVa} \\
\D_{s} \bar{u} &= -(2-\gamma_{D})\bar{u}, \hspace{0.8cm} \bar{u}(s=1) = u_{I}, \label{S4:eq.IVb} \\
\D_{s} \bar{z} &= -(2+\gamma_{\tau})\bar{z}, \hspace{0.925cm} \bar{z}(s=1) = z_{I}. \label{S4:eq.IVc}
\end{align}
where $ g_{I},u_{I} $ and $ z_{I} $ represents some initial values of invariant charges. The first equation can be readily integrated
\begin{align}
\ln s = \int_{g_{I}}^{\bar{g}} \frac{\dd x}{\beta_{g}(x)},
\end{align}
which defines $ \bar{g} $ implicitly as a function of $ g_{I} $. Eqs. \eqref{S4:eq.IVb} and \eqref{S4:eq.IVc} are then solved by changing the variables $ \D_{s} \rightarrow \beta_{g} \partial_{g} $, which gives
\begin{align}
\bar{u} = u_{I} s^{-2} \exp \left\{ \int_{g_{I}}^{\bar{g}} \dd x \frac{\gamma_{D}(x)}{\beta_{g}(x)} \right\}
\end{align}
and similarly for $ \bar{z} $. By knowing all invariant variables, the solution to RG equation is
\begin{align}
C(s,y,z,g) = C&(1,\bar{u},\bar{z},g) \nonumber \\
& \times \exp \left\{ \int_{g_{I}}^{\bar{g}} \dd x \frac{2\gamma_{\phi}(x) + \gamma_{D}(x)}{\beta_{g}(x)} \right\}
\end{align}
If the initial values of the running couplings $ g_{I} $ belong to a region of attraction of a IR stable fixed point $ g^{*} $, this fixed point will describe the universal scaling in the IR limit $ s \rightarrow 0, \ u \rightarrow 0, $ and $ z \rightarrow 0 $. All fixed points $ g^{*} $ are determined from solving the set of equations $ \beta_{g}|_{g\rightarrow g^{*}} = 0 $. Close to a fixed point, the RG flow equations behave as $ \beta_{g} \cong \Omega (g-g^{*}) $, where in the case of multi-charge theories $ \Omega $ is the stability matrix
\begin{align}
\Omega_{ij} = \partial_{g_{j}} \beta_{g_{i}}. \label{S4:eq.Omega}
\end{align}
For a fixed point to be IR stable, all eigenvalues $ e_{i} $ of \eqref{S4:eq.Omega} must be positive $ e_{i} > 0 $. 

Once the fixed points have been determined, the remaining integrals can be readily evaluated
\begin{align}
\int_{g_{I}}^{g^{*}} \dd x \frac{\gamma_{Q}(x)}{\beta_{g}(x)} = \gamma_{D}^{*} \ln s + c_{Q}(g_{I}), \label{S4:eq.gamma}
\end{align}
where $ \gamma_{Q}^{*} \equiv \gamma_{Q}(g^{*}) $ are the universal anomalous dimensions at the fixed point, and $ c_{Q}(g_{I}) $ are non-universal (regular) constants which depends on the initial point of the RG flow. The final forms of the invariant variables are
\begin{align}
\bar{u} \cong u_{n} s^{- 2 + \gamma_{D}^{*}}, \quad \bar{z} \cong z_{n} s^{-2-\gamma_{\tau}^{*}},
\end{align}
where from now on any quantity with a subscript `$ n $' denotes non-universal amplitude that contains the information about the initial conditions stored in $ c_{Q} $.

Putting all together, the solution to RG equation in the IR limit is 
\begin{align}
G^{(2,0)}(k,\omega,D,\tau,g,\mu)|_{\text{IR}} \nonumber \\
& \hspace{-2.5cm} \cong \frac{k^{2\Delta_{\phi}^{*} - d - \Delta_{D}^{*}}}{D_{n} \mu^{2\gamma_{\phi}^{*}-\gamma_{D}^{*}}} C \left( \frac{\omega \mu^{-\gamma_{D}^{*}}}{D_{n} k^{\Delta_{D}^{*}}}, \frac{\tau_{n} \mu^{\gamma_{\tau}^{*}}}{k^{\Delta_{\tau}^{*}}},g^{*} \right), \label{S4:eq.RGEq_sol}
\end{align}
where we have defined the total scaling dimensions as
\begin{align}
\Delta_{D}^{*} = 2 - \gamma_{D}^{*}, \quad \Delta_{Q}^{*} = d_{Q}^{k} + \gamma_{Q}^{*}, \quad Q \in \{ \phi,\phi',\tau \}.
\end{align}
The response function $ G^{(1,1)} $ can be obtained from \eqref{S4:eq.RGEq_sol} by performing the simple substitution $ 2\Delta_{\phi}^{*} \rightarrow \Delta_{\phi}^{*} + \Delta_{\phi'}^{*} $.

Eq. \eqref{S4:eq.RGEq_sol} represents the resummed perturbative expansion in $ g $; the appearance of the anomalous dimensions $ \gamma $ in the expressions $ (k/\mu)^{\gamma} $ is a result of resummed logarithmic divergences such as $ \ln(k/\mu) $, which appear in effective expansion parameters $ g^*\ln(k/\mu) $ of the renormalized model (see, e.g., \cite{Vasilev2004}). Therefore,   the fixed point values of the expansion parameters $ g^{*} $ must be small in order to ensure the consistency of the calculation (which can be systematically achieved only within the $ \varepsilon $-expansion).	 

Let us enclose this subsection with a further discussion on the r\^ole of the diffusion constant $ D $ \cite{Vasilev2004,AHKV05}. As we have mentioned before,  $ D $ does not represent a coupling constants (as it has non-zero frequency/momentum dimension), but nor it is a `mass' term as it is marginal in the context of $ \omega \sim k^{4} $ scaling. In order to understand its properties it is important to realize that the renormalization process replaces the bare diffusion constant with an effective diffusion constant that in the IR limit behaves as
\begin{align}
D_{0} \rightarrow \bar{D} \cong D_{n} (k/\mu)^{-\gamma_{D}^{*}}.
\end{align}
Depending on the sign of the anomalous dimension $ \gamma_{D}^{*} $ at the fixed point, the running diffusion constant either goes to zero or diverges in the limit $ (k/\mu) \rightarrow 0 $. However, it is not the above effective diffusion constant which is measured in the experiment, but its non-universal counter-part $ D_{n} $, which depends on the microscopic properties of the system. The anomalous scaling of $ \bar{D} $ determined by $ \gamma_{D}^{*} $ is then the origin of the modified dynamical exponent~$ \Delta_{D}^{*} $.

\subsection{IR behavior in $ d = 2 - \varepsilon $}

\begin{table*}
	\def\arraystretch{1.4}
	\centering
	\begin{tabular}{ c | c c c |c c c |c c c c }
		\hline \hline
		FP/$ g^{*}/\Delta $ & $\lambda^{*}$ & $\zeta^{*}$ & $ \eta^{*} $ & $e_{1}$ & $ e_{2} $ & $ e_{3} $ & $\Delta_{D}^{*}$ & $ \Delta_{\phi}^{*} $& $ \Delta_{\phi'}^{*} $ & $\Delta_{\tau}^{*}$
		\\  \hline
		FP0  & $ 0 $ & $ 0 $ & $ 0 $ & $ -\tfrac{1}{2}\varepsilon $ & $ -\tfrac{1}{2}\varepsilon $ & $ -\varepsilon $ & $ 4 $ & $-\tfrac{1}{2} \varepsilon$ & $2-\tfrac{1}{2} \varepsilon$ & $2$
		\\
		FPI  & $ a \sqrt{\vert\varepsilon\vert} $ & $ \frac{3a^{2}-4{\rm sgn}(\varepsilon)}{6a} \sqrt{\vert\varepsilon\vert} $ & $ 0 $ & $ \varepsilon $ & $ 0 $& $ -\tfrac{1}{3}\varepsilon $ & $ 4 - \tfrac{1}{3} \varepsilon $ & $ -\tfrac{1}{3}\varepsilon $ & $2-\tfrac{2}{3} \varepsilon$ & $2-\tfrac{1}{3}\varepsilon$
		\\
		FPII  & $ 0 $ & $ 0 $ & $ \tfrac{4}{3} \varepsilon $ & $ \tfrac{1}{6}\varepsilon $ & $ \frac{5}{18}\varepsilon $ & $ \varepsilon $ & $ 4 $ & $ -\tfrac{1}{2}\varepsilon $ & $2-\tfrac{1}{2} \varepsilon$ & $2+\tfrac{4}{9}\varepsilon$
		\\
		FPIII$ ^{\pm} $  & $ \pm \sqrt{\tfrac{4}{3}} \sqrt{\varepsilon} $ & $ \pm \sqrt{\tfrac{16}{3}} \sqrt{\varepsilon} $ & $ 4\varepsilon $ & $ -\varepsilon $ & $  \tfrac{1}{3}\varepsilon $ & $ \varepsilon $ & $ 4 + \varepsilon $ & $ -\varepsilon $ & $2$ & $2+\tfrac{7}{3}\varepsilon$
		\\ \hline \hline 
	\end{tabular}
	\caption{Table of fixed points with corresponding eigenvalues and critical exponents, calculated in the framework of $ \varepsilon $-expansion. The parameter $ a $ is a free parameter indicating that FPI is actually a line of fixed points. The $ \pm $ in FPIII indicates two fixed points, but they belong to the same universality class (they have the same critical exponents). Surprisingly, it seems that only FPII is stable. In FPI the negative eigenvalue corresponds to $ \eta $ direction, which means that the fixed point is unstable with respect to perturbations in $ \eta $ direction. Note that in FPII corrections to the scaling dimension are present only in the mass term (standard aspect of theories with quartic interaction). In 'standard' notation the exponents are $ \Delta_{\omega}^{*} = z, \Delta_{\phi}^{*} = - \chi $.}
	\label{tab:FPd2}
\end{table*} 

We will first analyse the critical scaling of the system close to $ d = 2 $, in which case all measurable quantities will be calculated in the context of $ \varepsilon $-expansion. The anomalous dimensions of all parameters and fields are calculated from the definitions \eqref{S4:eq.BetGgam} by keeping only the singular terms close to $ d = 2 $ in \eqref{S4:eq.Za}-\eqref{S4:eq.Zb}, and using \eqref{S4:eq.Zc}-\eqref{S4:eq.Zd}. The result is
\begin{align}
\gamma_{D} &= \tfrac{1}{4} \lambda  (\lambda - 2 \zeta ), \quad \gamma_{\phi} = - \gamma_{\phi'} = \tfrac{1}{8} \lambda  (\lambda - 2 \zeta), \\
\gamma_{\lambda} &= \tfrac{3}{8} \lambda  (2 \zeta -\lambda ) - \tfrac{1}{2} \eta, \quad \gamma_{\eta} = \tfrac{1}{2} \lambda  (2 \zeta -\lambda ) - \tfrac{3}{4} \eta, \\
\gamma_{\zeta} &= \tfrac{3}{8} \lambda  (2 \zeta -\lambda ) + \tfrac{1}{12} \eta  \left( 2\lambda/\zeta-7\right), \\
\gamma_{\tau} &= \tfrac{1}{4} \lambda
 (2\zeta-\lambda) + \tfrac{1}{3} \eta.
\end{align}
The corresponding RG flow equations are
\begin{align}
\beta_{\lambda} &= -\lambda (\tfrac{1}{2}\varepsilon + \gamma_{\lambda} ), \label{S4:eq.RGEq_1} \\
\beta_{\zeta} &= -\zeta (\tfrac{1}{2}\varepsilon + \gamma_{\zeta} ), \\
\beta_{\eta} &= -\eta (\varepsilon + \gamma_{\eta} ). \label{S4:eq.RGEq_2}
\end{align}
\begin{figure}
	\centering
	\includegraphics[width=\linewidth]{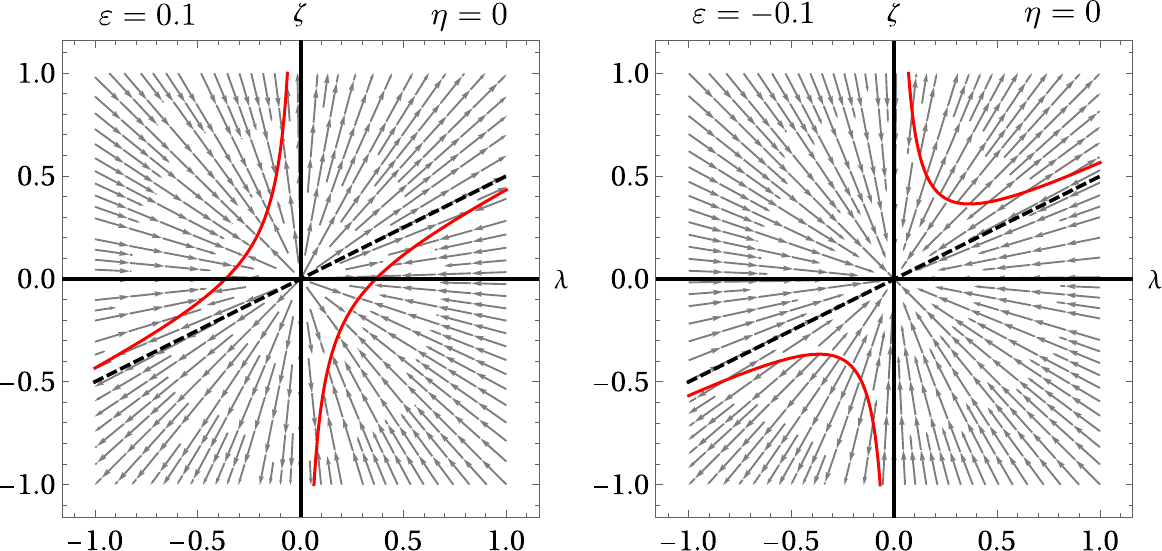}
	\caption{RG flow for CKPZ model in the absence of quartic interaction $ \eta = 0 $. The red line is the line of fixed points and the black dashed line is the critical line for the existence of the fixed points $ 2\zeta = \lambda $. Note that the flow slightly differs from \cite{Caballero2018} as the plots are done in the framework of $ \varepsilon $-expansion. } \label{fig:2d_eta0}
\end{figure}
The list of all fixed points with their stability eigenvalues, and critical exponents are shown in Tab \ref{tab:FPd2}. Besides the trivial (Gaussian) fixed point FP0 stable for $ \varepsilon < 0 $, we have found three universality classes:

\begin{itemize}
	\item FPI: The quartic interaction is irrelevant $ \eta^{*} = 0 $, which represents the CKPZ+ universality class reported in \cite{Caballero2018}. This fixed line is unstable with respect to perturbations in $ \eta $ direction. The corresponding RG flow in $ \eta = 0 $ plane is shown in Fig.~ \ref{fig:2d_eta0}. 
	
	\item FPII: In this case the CKPZ+ non-linearities are irrelevant, but the corresponding critical exponents attain no beyond mean-field corrections within the one-loop approximation. Nevertheless, the interface appears rougher than in the CKPZ+ class. This fixed point is stable for $ \varepsilon > 0 $, and unstable for $ \varepsilon < 0 $.
	
	\item FPIII: Most non-trivial fixed point where all non-linearities are important. This fixed point is, however, unstable for any $ \varepsilon $.
\end{itemize}
Let us further comment on the CKPZ+ universality class. An entire line of fixed points (determined by the free parameter $ a $ in Tab \ref{tab:FPd2}) appears because for $ \eta = 0 $ there are no divergent contributions to cubic non-linearities in the one loop approximation. This can be seen from Eqs. \eqref{S4:eq.Za}-\eqref{S4:eq.Zb}, where only the response function is renormalized in the absence of quartic interaction. Although the latter is known to be only an artefact of the one-loop approximation~\cite{Janssen1997}, higher-order terms in the $\varepsilon$ expansion do not affect lower-order terms, and the line of fixed points is preserved in all orders of perturbation theory. Moreover, the CKPZ fixed point ($ \lambda^{*} > 0 $ with $ \zeta^{*},\eta^{*} = 0 $) lies on the line of CKPZ+ fixed points, which can be seen from setting $ 3a^{2} = 4 $ in FPI. Such a fixed point does also exist in all orders of perturbation theory, since $ (\partial^{2}\phi')(\partial \phi)^{2} $ interaction alone does not generate corrections to $ (\partial_{i} \phi') (\partial_{i}\phi)(\partial^{2}\phi) $ nor $ (\partial_{i}\phi') (\partial_{i}\phi)(\partial\phi)^{2} $.

\begin{figure}
	\includegraphics[width=\linewidth]{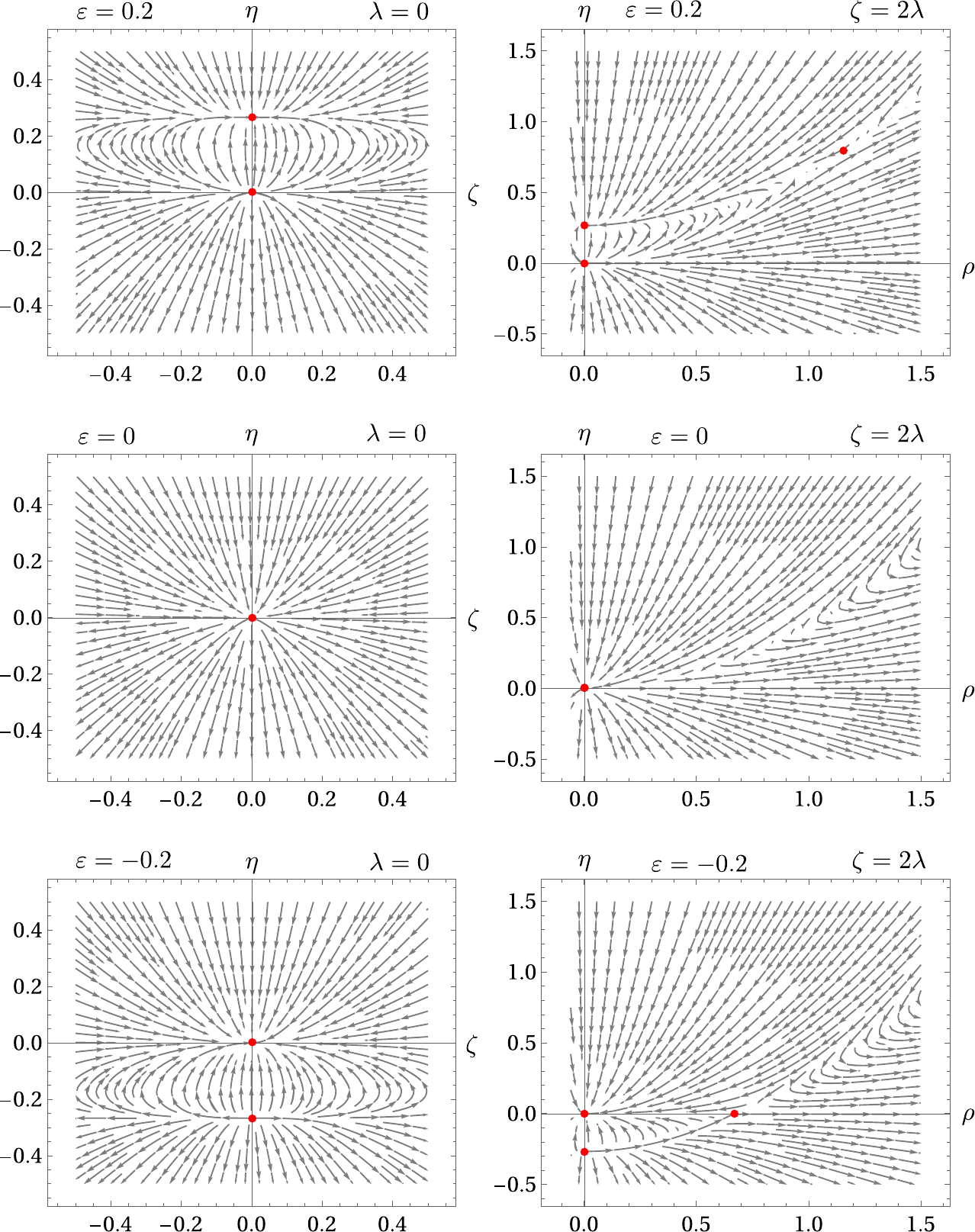}
	\caption{RG flow for the radial flows $ \lambda = 0 $ (left), and $ \lambda = 2 \zeta $ (right). Fixed points are denoted with a red dot (see discussion above). Note that there is an unstable region in the fixed plane $ \tan\theta = 2 $ for any $ \varepsilon $. This could be a sign of a strong coupling behavior.} \label{fig:2d_lambda0}
\end{figure}

In order to simply the analysis of the RG flow, we calculated the stability of the fixed point in the cylindrical coordinates (see Appendix \ref{APP:CC} for details)
\begin{align}
\lambda = \rho \cos \theta, \quad \zeta = \rho \sin \theta, \quad \eta = \eta.
\end{align}
In general, the RG flow is rotational arround the $ \lambda,\zeta = 0 $ line, but the angular component vanishes in two planes; The first is $ \lambda = 0 $, which represents an unstable plane, and the second one is $ \zeta = 2 \lambda $ which represents a stable plane. The RG flow in both of these planes is shown in Fig. \ref{fig:2d_lambda0}. The plot for $ \zeta = 2 \lambda $ shows that the only stable fixed point for $ \varepsilon > 0 $ is FPII, but only within a certain region of attraction. If the initial point of $ \rho $ is too large compare to $ \eta $, the RG flow goes to infinity. The size of this region changes for $ \zeta \neq 2\lambda $  and vanishes for $ \lambda = 0 $. The existence of this region could be in principle of a sign of a strong coupling behavior, similarly to what was suggested in \cite{Caballero2018} for $ \eta^{*} = 0 $. Similar ideas apply for $ \varepsilon = 0 $ and $ \varepsilon < 0 $ as well, where in the latter case an unstable fixed point at $ \eta = 0 $ appears.

\subsection{ IR behavior in $ d = 1 $} \label{S4:d1}

The special case $ d = 1 $ requires careful analysis. First of all, note that in $ d = 1 $ the Langevin equation \eqref{S1:eq.LE} becomes
\begin{align}
\partial_{t} \phi =&\ D (\tau - \partial_{x}^{2}) \partial_{x}^{2} \phi  - \tfrac{1}{2} \kappa \partial_{x}^{2} (\partial_{x}\phi)^{2} \nonumber \\
&\ + \tfrac{1}{3!} \eta \partial_{x} ((\partial_{x}\phi)(\partial_{x}\phi)^{2}) + f,
\end{align}
where $ \kappa = \lambda - \zeta/2 $ is a single coupling constant. This means, that when calculating perturbative corrections in general $ d $, the number of coupling constants must reduce from three to two after setting $ d = 1 $.

It is important to note that the above is merely a property of the perturbation theory, and says nothing about the RG procedure. As we have mentioned before, the RG analysis can give reliable results \emph{only} within the framework of $ \varepsilon $-expansion. The latter however, prevents us from introducing a single coupling constant. Hence, we will derive the RG flow equations at fixed $ d = 1 $, but one must keep in mind the possible inaccuracy of the results.

Following the same procedure as before, the renormalised response functional \eqref{3:eq.RSF} becomes 
\begin{align}
\s[\varphi] =&\  \tfrac{1}{2} Z_{1} D \phi'\partial^{2}
\phi' + \phi' \{ \partial_{t} + D (Z_{2} \partial^{4} - Z_{6} \tau \partial^{2} )  \} \phi \nonumber \\
&\ + \tfrac{1}{2} D Z_{3,4} \mu^{\varepsilon/2} \kappa (\partial^{2}\phi') (\partial \phi)^{2} \nonumber \\
&\ + \tfrac{1}{3!} D Z_{5} \mu^{\varepsilon} \eta (\partial_{i} \phi')(\partial_{i}\phi)(\partial\phi)^{2},
\end{align}
where a coupling constant $ \kappa_{0} = \lambda_{0} - \zeta_{0}/2 $ was introduced. The multiplicative renormalization of the above model is still ensured (see Sec. \ref{S3:A}), where the new coupling renormalizes as
\begin{align}
\kappa_{0} = \mu^{\varepsilon/2} \kappa Z_{\kappa} . \quad  \kappa Z_{\kappa} = (\lambda Z_{\lambda} - \zeta Z_{\zeta}/2) |_{d = 1},
\end{align}
which gives
\begin{align}
Z_{\kappa} = 1 + \tfrac{15}{32} \kappa^{2} + \tfrac{9}{16} \eta.
\end{align}
The rest of the calculation follows an analogous procedure to the $ d = 2 - \varepsilon $ calculation, and the resulting RG flow equations for fixed $ d = 1 $ case are
\begin{align}
\beta_{\kappa} &= - \kappa \big(\tfrac{1}{2} - \tfrac{1}{8} ( 5 \kappa^{2} + 6 \eta) \big), \\ 
\beta_{\eta} &= - \eta \big(1 - \tfrac{3}{32} ( 5 \kappa^{2} + 6 \eta) \big).
\end{align}
The list of fixed points, with their stability eigenvalues and critical exponents are shown in Tab. \ref{S4:Tab.FP}.

It is interesting to see that the most non-trivial fixed point FPIII disappears in $ d = 1 $, and the only existing fixed points are FPI$ |_{d=1} $ and FPII$ |_{d=1} $ analogous to those in $ d = 2 - \varepsilon $ case. Moreover, almost all critical scaling dimensions in $ d = 1 $ are identical to the ones in $ d = 2 - \varepsilon $ with $ \varepsilon = 1 $. The only exception is $ \Delta_{\tau}^{*} $ for FPII, where the difference is $ 1/9 $.

\begin{figure}[t!]
	\centering
	\includegraphics[width=\linewidth]{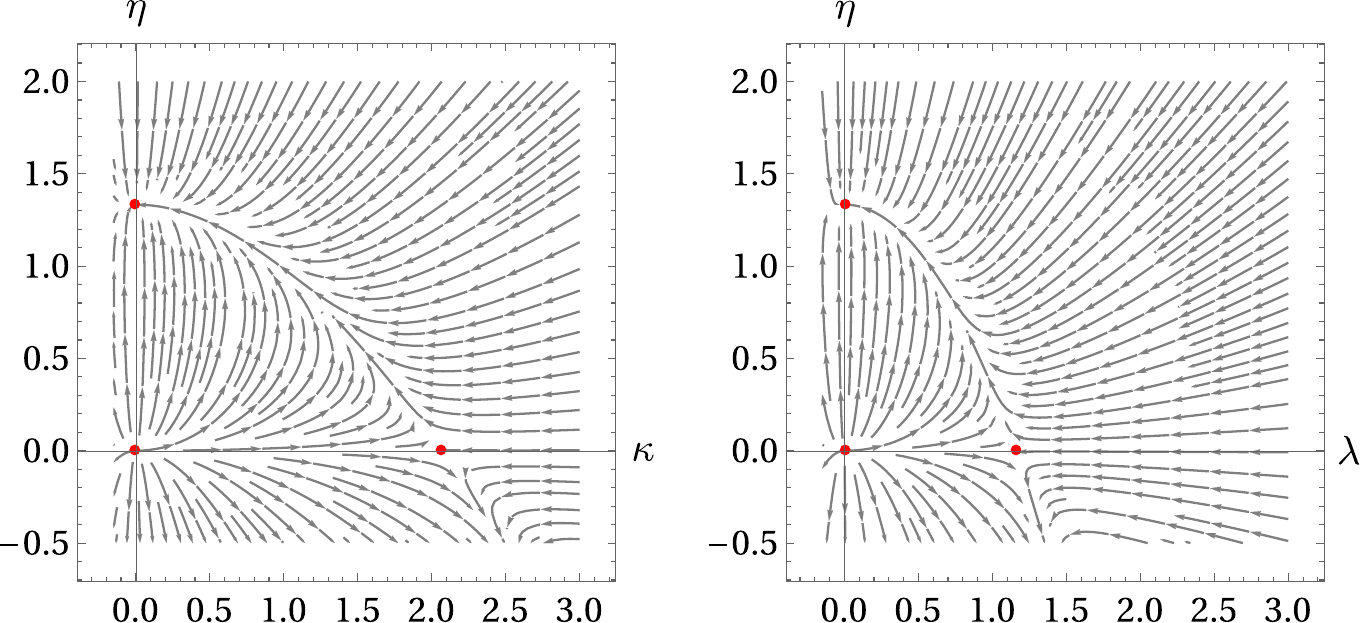}
	\caption{The RG flow in the $\eta$-$\kappa$ plane calculated at fixed $ d = 1 $ (left), and analytical extrapolation from $ d = 2 $ to $ d = 1 $ for $ \zeta = 0 $ (right). 
	} \label{S4:Tab.d1RG}
\end{figure}

The RG flow for $ d = 1 $ is plotted in Fig. \ref{S4:Tab.d1RG}. Similarly to the $ d = 2 - \varepsilon $ case, the only stable fixed point for $ \eta > 0 $ is FPII. This means, that the general conserved surface growth in $ d = 1 $ (for small $ \tau $) is probably described by the model \eqref{S1:eq.LE-IR}. For comparison, we have also plotted the RG flow in $ \zeta = 0 $ plane within the $ \varepsilon $-expansion framework for $ \varepsilon = 1 $. The flows are qualitatively the same and they give the same IR fixed point.

\begin{table}[t!]
	\def\arraystretch{1.2}
	\centering
	\begin{tabular}{ c| c c | c c | c c c c }
		 \hline  \hline 
		FP/$ g^{*}/\Delta $ & $(\kappa^{*})^{2}$ & $ \eta^{*} $ & $e_{\kappa}$ & $ e_{\eta} $ & $\Delta_{D}^{*}$ & $ \Delta_{\phi}^{*} $ & $ \Delta_{\phi'}^{*} $ & $ \Delta_{\tau}^{*} $
		\\  \hline 
		FP0$ |_{d=1} $  & $ 0 $ & $ 0 $ & $ -1/2 $ & $ -1 $ & $ 4 $ & $-1/2$ & $3/2$ & $2$
		\\  
		FPI$ |_{d=1} $  & $ 16/15 $ & $ 0 $ & $ 1 $ & $ -1/3 $& $ 11/3 $ & $ -1/3 $ & $ 4/3 $ & $ 5/3 $
		\\  
		FPII$ |_{d=1} $ & $ 0 $ & $ 4/3 $ & $ 1/4 $ & $ 1 $ & $ 4 $ & $ -1/2 $ & $ 3/2 $ & $ 7/3 $
		\\ \hline  \hline 
	\end{tabular}
	\caption{Table of fixed points, eigenvalues and critical exponents. The fixed point that remains the same is the same as in the $ d = 2 - \varepsilon $ expansion in the previous section. In `standard' notation the exponents are $ \Delta_{\omega}^{*} = z, \Delta_{\phi}^{*} = - \chi $.} \label{S4:Tab.FP}
\end{table}

\subsection{The diffusive regime ($ d_{\omega} = 2 $)} \label{S6:D} 

All results derived so far where done in the limit $ k \sim \omega^{4} \rightarrow 0 $. In such case, the parameter $ \tau $ had a dimension of mass, hence it grew under the RG transformation unless $ \tau \ll 1 $. This means, that away from the critical point the system can be described by a different fixed point with a different dispersion law. In order to derive the scaling properties in this regime, one must study the limit $ \omega \sim k^{2} \rightarrow 0 $ instead. Here, the $ \partial^{4} $ terms is IR irrelevant, and the response functional \eqref{S3:eq.RF} becomes
\begin{align}
\s_{0}[\varphi] =&\ -\tfrac{1}{2} \tau_{0} (\partial_{i}
\phi')^{2} + \phi' \{ \partial_{t} - \tau_{0} \partial^{2} \} \phi \nonumber \\
&\ + \tau_{0} \big[ \tfrac{1}{2} \lambda_{0} (\partial^{2}\phi') (\partial\phi)^{2} + \tfrac{1}{2} \zeta_{0} ( \partial_{i} \phi' )(\partial_{i} \phi) (\partial^{2} \phi) \nonumber \\
&\ + \tfrac{1}{3!} \eta_{0} ( \partial_{i} \phi' )(\partial_{i} \phi) (\partial \phi)^{2}  \big],  \label{S3:eq.RFdo2}
\end{align}
where we have rescalled the amplitude of the noise and all interactions with the parameter $ \tau_{0} $.

The renormalization process now proceeds in the analogous way to what was done in Sec. \ref{S3:}. The canonical dimensions in the present limit are shown in Tab.~\ref{tab:canon_CKPZdo2}. Here, the upper critical dimension for all non-linearities turns out to be $ d_{c} < 0 $, which means that any realistic scenario is described by the Gaussian fixed point. We conclude that the dynamics of the conserved surface roughening away from the critical point $ \tau \gtrsim 1 $ is driven by the following simple stochastic field equation
\begin{align}
\partial_{t} \phi =&\  \partial_{i} ( \tau \partial_{i} \phi + \sqrt{\tau} f_{i} ), \label{S4:eq.EWCN}
\end{align}
with $ f_{i} $ being the Gaussian random noise. The Eq. \eqref{S4:eq.EWCN} which is nothing else than just Edward-Willkinson equation with conserved noise \cite{barabasi1995}. The correlation and response functions then attain the following form
\begin{align}
\langle \phi \phi \rangle = \frac{\tau  k^{2}}{\omega^{2} + \tau ^{2}k^{4}}, \quad \langle \phi \phi' \rangle = \frac{1 }{-i\omega + \tau k^{2}}.
\end{align}
with no corrections at all.

\begin{table}
	\def\arraystretch{1.3}
	\centering
	\begin{tabular}{ c | c c c c }
		\hline \hline
		$Q$ & $ \phi',\phi $ & $ \tau_{0},\tau $ & $ \lambda_{0},\zeta_{0} $ & $ \eta_{0} $
		\\  \hline
		$d_Q^k$ & $ d/2 $ & $ 2 $ & $ -(4+3d)/2 $ & $ -2(1+d) $
		\\  
		$d^\omega_Q$ & $ 0 $ & $ -1 $ & $ 0 $ & $ 0 $
		\\  
		$d_Q$ & $ d/2 $ & $ 0 $ & $ -(4+3d)/2 $ & $ -2(1+d) $
		\\ \hline \hline 
	\end{tabular}
	\caption{Canonical dimensions of the bare fields and bare parameters, for the model \eqref{S3:eq.RFdo2}, with $ d_{Q} = d_{Q}^{k} + 2 d_{Q}^{\omega} $.}
	\label{tab:canon_CKPZdo2}
\end{table}

\subsection{Comparison with earlier works} \label{S4.E}

In a number of recent works \cite{Villain1991,Tang1991,Lai1991,Haselwandter2007_PRL,Haselwandter2007,Haselwandter2008}, authors considered the model \eqref{S1:eq.LE} without the `CKPZ+' term $ \partial_{i}(\partial_{i}\phi\partial^{2}\phi) $, but with the following noise correlator
\begin{align}
- D \partial^{2} \rightarrow D_{0} - D_{2} \partial^{2} - D_{4} \partial^{4}. \label{S4:eq.compar_Noise}
\end{align}
Most of the RG calculations of such model have been carried out within the context of Wilson's RG approach in fixed $ d $. In this last section we will discuss the outcome of these works and compare them with the present results.

The starting point for any RG calculation should always be dimensional analysis. In dynamical systems one must also chose the scale connection $ \omega \sim k^{d_{\omega}} $, where two physically relevant regimes exist for this model: $ d_{\omega} = 2 $ and $ 4 $. The upper critical dimension $ d_{c} $ is then invariably determined by the leading IR terms, and higher order terms are irrelevant which can be nicely understood from the Wilson's RG picture \cite{Wilson1974}. If one choses to keep the non-conserved noise $ D_{0} $ as the leading IR term in \eqref{S4:eq.compar_Noise}, then both $ D_{2}\partial^{2} $ and $ D_{4}\partial^{4} $ are IR irrelevant from the perturbative RG sense (as they contain higher powers of $ \partial $), and hence can be discarded from the very beginning of the calculation.
This regime was analysed in  \cite{Tang1991,Lai1991,Haselwandter2007,Haselwandter2007_PRL,Haselwandter2008}, where the upper critical dimension is $ d_{c} = 4 $. Note, that all of the results derived for this regime are, strictly speaking, valid only close to $ d_{c} $, although this important point is often absent in the literature. 

It is normally claimed that $ D_{4}\partial^4 $ term must be added in order to capture the leading-order correction to the noise correlator \cite{Tang1991,Haselwandter2007,Haselwandter2008}. Although  there is in principle nothing incorrect about the latter statement (which also holds to all orders of perturbation theory due to \eqref{S3:eq.diagsO45}), $ D_{4}\partial^{4} $ term is IR irrelevant in the presence of $D_0$, and does \emph{not} need to be added in order to perform a consistent Wilson's RG analysis~\cite{Wilson1974}. In the absence of $ D_{0} $ on the other hand (or in the case where $ D_{0} \ll 1 $), the leading IR term is the conserved noise $ -D_{2}\partial^{2} $. This limit was analyzed in the current work, as we demanded the model to obey the mass conservation law \eqref{S2:B.ContEq}. Different leading term in the noise spectrum implied different upper critical dimension $ d_{c} = 2 $, but as long as the conservative noise is non-vanishing, $ D_{4}\partial^{4} $ remains irrelevant and cannot influence the IR scaling. 

Let us now proceed with a discussion about technical issues arsing from Feynman diagram calculations in fixed dimension. As we have mentioned in Sec. \ref{S3:D} in field theory the UV cut-off does not only simply limits the integration domain to $ k < \Lambda $, but it regularizes every propagator in the perturbative expansion. The similar situation occurs in the Wilson's RG approach, where the propagators should have the following form
\begin{align}
\langle \phi \phi' \rangle_{0}|_{\text{reg}}^{\text{W}} &= \frac{\theta(\Lambda - k) - \theta(\Lambda/b - k)}{-i\omega + D_{0}k^{2}(k^{2} + \tau_{0})}, \label{4E.Wp1} \\
\langle \phi \phi \rangle_{0}|_{\text{reg}}^{\text{W}} &= \frac{\theta(\Lambda - k) - \theta(\Lambda/b - k)}{\omega^{2} + D_{0}^{2}k^{4}(k^{2} + \tau_{0})^{2}}, \label{4E.Wp2}
\end{align}
which represents a straightforward generalization of the regularized static propagators \cite{Wilson1974}.
Hence, the momentum integrals are not necessary evaluated over the simple spherical `shell' $ \Lambda/b < k < \Lambda $, but rather over the integration domain that is determined by the propagators \eqref{4E.Wp1}-\eqref{4E.Wp2}. Without including the regulating functions in the above form, the results of the diagrams are not invariant with respect to the shift of internal momentum. For example, the integral Eq. (A10) in \cite{Haselwandter2007} with a shift $ \qq \rightarrow \qq - \kk/2 $ gives (in their notation)
\begin{align}
\Phi_{1} = - k^{4}\frac{K_{d}}{d} \frac{2\lambda_{22}^{2}(D_{0} + D_{2} \Lambda^{2} + D_{4} q^{4}) \Lambda^{d}}{(\nu_{2} + \Lambda^{2} \nu_{4})^{3}} \dd l,
\end{align} 
instead of their (A14). Note, that these two results match only for $ d = 4, \ D_{2,4} \rightarrow 0, \ \nu_{2} \rightarrow 0 $ or $ d = 2, \ D_{0,4} \rightarrow 0, \ \nu_{2} \rightarrow 0 $. In order to perform a consistent series expansion in terms of external momenta, the cut-off functions must be expanded as well. For practical calculations it is more convenient to work with a smooth cut-off, for example $ \theta(\Lambda-q) =  \mathrm{e}^{-q^{2}/\Lambda^{2}} $ \cite{Zinn}. For the symmetric split of Eq. (A10) from \cite{Haselwandter2007}, this can be achieved by inclusion of the following terms into the integrand (in their notation again)
\begin{align}
& \left( {\rm e}^{-\frac{(\kk/2-\qq)^{2}}{\Lambda^{2}}} - {\rm e}^{-\frac{(\kk/2-\qq)^{2}}{\Lambda^{2}/b^{2}}} \right) \left( {\rm e}^{-\frac{(\kk/2 + \qq)^{2}}{\Lambda^{2}}} - {\rm e}^{-\frac{(\kk/2 + \qq)^{2}}{\Lambda^{2}/b^{2}}} \right) \nonumber \\
& \hspace{1.5cm} = \left( {\rm e}^{-\frac{\qq^{2}}{\Lambda^{2}}} - {\rm e}^{-\frac{\qq^{2}}{\Lambda^{2}/b^{2}}} \right)^{2} + \mathcal{O}(k^{2}), \label{S4:E.regExp}
\end{align}
while keeping the integration domain $ |\qq| \in (0,\infty) $. The first term containing decaying exponentials represents a `smooth' form of the momentum shell $ \Lambda/b < k < \Lambda $, while the latter terms contains the corrections calculated in the series of $ k $.

The result \eqref{S4:E.regExp} shows, that the next-to-leading order contribution coming from expanding the regulating functions is actually of order $ \mathcal{O}(k^{2}) $. As the integrand Eq. (A10) in \cite{Haselwandter2007} is already of order $ k^{3} $ due to the structure of external vertexes, no marginal contributions arise from the series expansion of the regulating functions, and one can safely set $ k = 0 $ in \eqref{S4:E.regExp}. Note, that this is \emph{only} the case for the symmetric split of the external momenta: the next-to-leading contributions coming from an asymmetric split would be of order $ k $ which will have to be included in the calculation. We would like to point out that this is also the reason why Wilson's RG calculation of standard KPZ equation in fixed $ d $ presented in \cite{FreyTauber94} matches with their field-theoretic RG results carried out in dimensional renormalization.

\section{Conclusion}

	In this work, we have been studying scaling properties of a general equation describing conserved surface roughening. Starting from a conservation law, we have constructed a stochastic field model by including all terms that are allowed by the symmetry of the system: $ i) $ standard CKPZ nonlinearity, $ ii) $ CKPZ+ nonlinearity that is distinguishable from the former only in $ d > 1 $, and $ iii) $ cubic nonlinearity not analysed before in the context of conserved surface roughening, to our best knowledge.

In general, the system shows two fundamentally different scaling regimes, where the mean-field dynamical critical exponent is either $ z = 2 $, or $ z = 4 $. The first regime appears when the gravitational forces responsible for the surface particle movement dominates over the curvature effects $ \tau / D \gg 1 $. Here, we have shown that all nonlinearities are irrelevant at any spatial dimension, so that the universal scaling is determined by Edward-Wilkinson equation with conservative noise \eqref{S4:eq.EWCN}. The $ z = 4 $ regime appears when the particles are driven mainly by the surface curvature $ \tau/D \ll 1 $. In this case both quadratic and cubic nonlinearities turned out to be marginal at the upper critical dimension $ d_{c} = 2 $, so that corrections to scaling exponents must be taken into account.

Using standard tools, the stochastic field model was mapped into the response functional formalism, which allowed us to study the  large-scale and long-time properties using the well known methods of field-theoretic renormalization group. The structure of the perturbation theory was discussed, followed by the discussion about the renormalization. We have shown that in the limit $ \omega \sim k^{4} \rightarrow 0 $, the model is renormalizable at the upper critical dimension $ d_{c} = 2 $ as long as it has a form of gradient expansion in terms of the height field $ \partial \phi $. Further analysis of the divergent structures revealed that the system can be decomposed into two sub-class of models: model with cubic and quartic interactions only (or, in the context of Langevin formulation to model with quadratic and cubic nonlinearity).

In order to avoid difficulties arising from UV cut-off regularization, the UV renormalization was carried out in the dimensional renormalization. As the two quadratic non-linearities are distinguishable only for $ d > 1 $, additional relation between the renormalization constants appears in $ d = 1 $. Normalization point scheme was used in order to capture this relation. The RG equation was then derived, which solution revealed the IR scaling properties.

The critical properties in the $ \omega \sim k^{4} \rightarrow 0 $ limit have been derived by the means of both $ \varepsilon = 2 - d $ expansion and fixed dimension $ d = 1 $. These approaches are in the qualitative agreement for $ \varepsilon = 1 $. In the absence of the quartic interaction $ \eta = 0 $, the system remains in the CKPZ+ universality class, which properties were reported before. We have shown that this results also holds to all orders of the perturbation theory. Below the upper critical dimension $ \varepsilon > 0 $, and for positive values of $ \eta $, the RG flow either shows a run-away solution or it flows to a fixed point where only the quartic interaction is non-zero $ \eta^{*} > 0, \  \lambda^{*}, \zeta^{*} = 0 $. The corrections to mean-field critical exponents are absent in the one-loop approximation (the only exception is $ \tau $), but should be present starting from two-loops. The most non-trivial fixed point with all interactions being relevant $ \eta^{*}, \lambda^{*}, \zeta^{*} > 0 $ is unstable at any $ \varepsilon $. Intriguingly, the presence of this fixed point is absent in the fixed $ d = 1 $ calculations, which demonstrates the difference between the `fixed $ d $' and the $\varepsilon$-expansion schemes. 

We have finalized the paper with discussing the relation of our results to previous works. Moreover, we have also pointed out the commonly overlooked technical difficulties arising in the Wilson's RG approach, which are especially important for the fixed $ d $ calculations.

\section{Acknowledgment}

The authors would like to thank Francesco Cagnetta, Cesare Nardini, and Calum Milloy for many illuminating discussions. V.\v{S}. acknowledges studentship funding from EPSRC grant no. EP/L015110/1.

\appendix

\section{Degree of divergence} \label{APP:DoD}

In this appendix we will derive Eq. \eqref{S3:eq.dG} based on topological properties of the model. The UV exponent of any Feynman diagram is \cite{Vasilev2004,Zinn,Amit2005}
\begin{align}
d[I_{nN}] = (d + 4) L + \sum_{i} d_{I_{i}} I_{i} + \sum_{j} d_{V_{j}} V_{j}, \label{Sec2:eq.1}
\end{align}
where $ L $ stands for number of loop integrals, $ I_{i}, V_{j} $ are number of $ i $-th internal propagators and number of $ j $-th internal vertexes. As we are interested in the leading UV behavior of Feynman diagrams, the dimensions of propagators $ d_{I_{i}} $ are determined by keeping only the leading UV terms in the denominators. For the present model, these are $ d_{I_{\phi\phi}} = -6 $, and $ d_{I_{\phi\phi'}} = - 4 $. All vertexes have dimension $ 4 $. The topological properties of the perturbation theory which relates these quantities can be analysed in the following way. The field $ \phi $ can be attached twice to $ V_{\phi'\phi\phi} $ vertex and three times to $ V_{\phi'\phi\phi\phi} $ vertex. This number must be equal to the number of fields on the internal propagators and external fields: $ 2 $ times on the $ \langle \phi\phi \rangle $ propagator, once on the $ \langle \phi \phi' \rangle $, and remaining fields are attached to external points. Summing up, we have
\begin{align}
\phi: \quad  2V_{\phi'\phi\phi} + 3 V_{\phi'\phi\phi\phi} &= 2I_{\phi\phi} + I_{\phi'\phi} + N_{\phi}, \label{Sec2:eq.2} \\
\phi': \quad \quad V_{\phi'\phi\phi} + V_{\phi'\phi\phi\phi} &= I_{\phi'\phi} + N_{\phi'}, \label{Sec2:eq.3}
\end{align}
where $ N_{\phi} $ and $ N_{\phi'} $ are total number of external $ \phi $ and $ \phi' $ fields, and the second relation was obtained in the analogous way for the field $ \phi' $. The number of loops is $ L = I - (V - 1) $, where $ I,V $ are total number of internal propagators and internal vertexes \cite{Vasilev2004,Zinn,Amit2005}. Eqs. \eqref{Sec2:eq.2}-\eqref{Sec2:eq.3} allows us to write the degree of divergence \eqref{Sec2:eq.1} only as a function of number of external points and vertexes
\begin{align}
d[I_{nN}] =&\ d + 4 - \tfrac{1}{2} (d+2) N_{\phi'} - \tfrac{1}{2} (d-2) N_{\phi} \nonumber \\
&\ - \tfrac{1}{2} (2-d) V_{\phi'\phi\phi} - (2-d) V_{\phi'\phi\phi\phi}. 
\end{align}
which is exactly Eq. \eqref{S3:eq.dG} in the main text.

\section{Feynman diagrams} \label{APP:FD}

All calculations have been carried out in the Fourier space where the following convention has been used
\begin{align}
F(\xx,t) &= \int \frac{\dd \kk \dd \omega}{(2\pi)^{d+1}} \ F(\kk,\omega) {\rm e}^{i(\kk.\xx-\omega t)} \\
F(\kk,\omega) &= \int \dd \xx \dd t \ F(\xx,t) {\rm e}^{-i(\kk.\xx-\omega t)}
\end{align}
The calculation of all diagrams have been done with the $ \tau $ IR cut-off in all propagators, and using the well-known formula \cite{Vasilev2004,Zinn}
\begin{align}
\int {\dd \kk \over (2\pi)^{d} } \frac{ k^{2\beta}}{(k^{2} + \tau)^{\alpha}} = \frac{\G[d/2+\beta]\G[\alpha - \beta - d/2]}{(4\pi)^{d/2}\G[d/2]\G[\alpha]} \tau^{d/2+\beta-\alpha} .\label{App:FDform}
\end{align}
The repeating constant $ A_{d} = \G[3-d/2]/(4\pi)^{d/2} $ finite at $ d = 2 $ has been absorbed into the coupling constants
\begin{align}
A_{d} \times \{ \lambda^{2},\lambda\zeta,\zeta^{2},\eta \} \rightarrow \{ \lambda^{2},\lambda\zeta,\zeta^{2},\eta \}. \label{eq:diff}
\end{align}

\subsection{Two-point diagrams}

The only non-zero two point diagram is the sun-set diagram contributing to response function
\begin{align}
\raisebox{-0.34cm}{\includegraphics[height=0.9cm]{psp_1.pdf}}.
\end{align}
Its explicit form is found using Feynman rules \eqref{S2:eq.p1}-\eqref{S2:eq.v2}
\begin{align}
\int {\dd \kk \dd \omega \over (2\pi)^{d+1}}\, \frac{D_{0} k^{2} \mathcal{V}_{\pp,\kk-\pp,-\kk} \mathcal{V}_{\pp-\kk,-\pp,\kk}}{[\omega^{2} + \epsilon_{\kk}^{2}][-i(\Omega-\omega)+\epsilon_{\pp-\kk}]} ,
\end{align}
where
$ \epsilon_{\kk} = D_{0} k^{2}(k^{2} + \tau_{0}) $. The frequency integral is readily performed using the residue theorem. The resulting expression is expanded in terms of the external momenta $ \pp $, while keeping only terms up to $ p^{4} $. The final momentum integral is carried out within the help of the following formulas
\begin{align}
\int{\dd \kk \over (2\pi)^{d}}
k_{i} k_{j} f(k^{2}) =&\ \frac{\delta_{ij}}{d} 
 \int {\dd \kk \over (2\pi)^{d}}
 k^{2} f(k^{2}), \\
\int {\dd \kk \over (2\pi)^{d}}
k_{i}k_{j}k_{k}k_{l}  f(k^{2}) =&\ \frac{\delta_{ij}\delta_{kl} + (2 \ \text{perm})}{d(d+2)} 
\\\times &\int {\dd \kk \over (2\pi)^{d}}
k^{4} f(k^{2}),
\end{align}
and Eq. \eqref{App:FDform}. All integrals with odd number of $ k_{i} $ components vanish due to reflection invariance of the measure. The final result is
\begin{align}
& \raisebox{-0.34cm}{\includegraphics[height=0.9cm]{psp_1.pdf}} = (-D p^4)  \times \Bigg( \frac{\lambda  (\lambda - 2 \zeta )}{4 (2-d)} \nonumber \\
& \hspace{0.3cm} +  \frac{(\zeta + 2 \lambda ) ((7 d+8) \zeta + 2(d+2) \lambda )}{64 (d+2)} \Bigg) \left( \frac{\mu^{2}}{\tau} \right)^{\varepsilon/2}, \label{app:diag1}
\end{align}
where we have replaced bare quantities with their renormalised counter-parts, which is a valid step in the leading order of the perturbation theory.

The correction to mass is calculated in the analogous way
\begin{align}
\frac{1}{2} \ \partial_{\tau} \bigg( \raisebox{-0.17cm}{\includegraphics[height=0.65cm]{psp_2.pdf}} \bigg) = D p^{2} \times \frac{(2+d)\eta}{12(2-d)} \left( \frac{\mu^{2}}{\tau} \right)^{\varepsilon/2}.
\end{align}

For completeness, we mention that the correction to the noise correlator vanishes at one-loop approximation due to the structure of the perturbation theory described in Sec. \ref{S3:B}. 
\begin{align}
\frac{1}{2} \ \raisebox{-0.34cm}{\includegraphics[height=0.9cm]{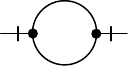}} \ = 0.
\end{align}

\subsection{Three-point diagrams}

In general, corrections to the three-point vertex function contain $ 2 $ external momenta. Due to the $ p\leftrightarrow r $ symmetry, they have the following form
\begin{align}
A (\pp.\rr)^{2} + B \pp.\rr ( p^{2} + r^{2}) + C p^{2} r^{2}. \label{App:eq.vertex2} 
\end{align}
In order for this expression to have the form similar to Eq. \eqref{S4:eq.corr3Point}, we rewrite it as
\begin{align}
\tfrac{1}{2}
& A [2(\pp.\rr)^{2} + \pp.\rr(p^{2} + r^{2})] \nonumber \\ 
& \hspace{1cm} + \tfrac{1}{2}[2Cp^{2}r^{2} + (2B-A)\pp.\rr(p^{2}+r^{2})],
\end{align}
which implies that $ 2B-A = C $ must hold. The one loop normalization constants are then equal to
\begin{align}
Z_{3} = 1 + \tfrac{1}{2 D\lambda} A, \quad Z_{4} = 1 - \tfrac{1}{D\zeta} C .
\end{align} 
The calculation of the Feynman diagrams goes in the analogous way as in the case of sun-set diagram. The results of the diagrams are listed below
\begin{align}
\raisebox{-0.5cm}{\includegraphics[height=1.2cm]{pspp_1.pdf}} &= \frac{D \mu^{\varepsilon/2} (\zeta - 2\lambda )^2}{128 (2-d)} \left( \frac{\mu^{2}}{\tau} \right)^{\varepsilon/2} \times \nonumber \\ 
&\ \hspace{-1.2cm} \times \left( X (\pp.\rr)^{2} +  Y (\pp.\rr)(p^{2} + r^{2}) + Z p^{2} r^{2} \right), \ \\
X&= 2 ((2-d) \zeta - 2(d+2) \lambda ) \lambda, \\
Y&= (4-d) \zeta - 2(2+d) \lambda, \\
Z&= 4 \zeta. \\
\raisebox{-0.5cm}{\includegraphics[height=1.2cm]{pspp_1.pdf}} &= - 2 \ \raisebox{-0.48cm}{\includegraphics[height=1.2cm]{pspp_2.pdf}}.
\end{align}
The above cancellation of the one-loop corrections to CKPZ equation without the quartic vertex is most probably the artefact of the one-loop approximation \cite{Janssen1997}. Inclusion of the quartic vertex leads to the following two diagrams.
\begin{align}
\raisebox{-0.34cm}{\includegraphics[height=0.9cm]{pspp_3.pdf}} &= \frac{D\mu^{\varepsilon/2}\eta }{24 (4-d^{2})} \left( \frac{\mu^{2}}{\tau} \right)^{\varepsilon/2} \times \nonumber \\
&\ \hspace{-1.2cm} \times \left( X (\pp.\rr)^{2} + Y (\pp.\rr)(p^{2} + r^{2}) + Z p^{2} r^{2} \right),  \\
X&= 2 \left(\left(d^2+2 d-8\right) \zeta +2(d+2)^2 \lambda \right), \\
Y&= \left(d^2+2 d-12\right) \zeta + 2(d+2)^2 \lambda, \\
Z&= -8 \zeta .
\end{align}
For the last diagram, two possible permutations of the external momenta must be considered
\begin{align}
2 \ \raisebox{-0.34cm}{\includegraphics[height=0.9cm]{pspp_4.pdf}} =&\ \ \raisebox{-0.34cm}{\includegraphics[height=0.9cm]{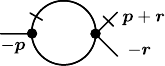}} \nonumber \\
&\ + \raisebox{-0.34cm}{\includegraphics[height=0.9cm]{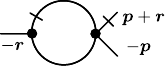}}.
\end{align}
The final result is
\begin{align}
2 \ \raisebox{-0.34cm}{\includegraphics[height=0.9cm]{pspp_4.pdf}} &= \frac{D\mu^{\varepsilon/2} \eta }{96 (4-d^{2})} \left( \frac{\mu^{2}}{\tau} \right)^{\varepsilon/2} \times \nonumber \\
&\hspace{-1.4cm} \times \left( X (\pp.\rr)^{2} +  Y (\pp.\rr)(p^{2} + r^{2}) + Z p^{2} r^{2} \right), \\
& \hspace{-1cm} X = 8 ((2-d) \zeta + 2(d+6) \lambda ), \\
& \hspace{-1cm} Y = -(d+4) ((7 d+2) \zeta +2(d-10) \lambda ), \\
& \hspace{-1cm} Z = -2 \big(\left(7 d^2+26 d+16\right) \zeta \nonumber \\
& + 2\left(d^2-2 d-16\right) \lambda \big).
\end{align}

\subsection{Four-point diagrams}

For this vertex function, every diagram must be calculated multiple times with different permutation of external momenta. The number of permutations is determined by the multiplicity factor in the front of the diagram. The results are
\begin{align}
3 \ \raisebox{-0.34cm}{\includegraphics[height=0.9cm]{psppp_1.pdf}} &= F_{\pp,\qq,\rr} \frac{\left(d^2+6 d+20\right) \eta }{12 (d^{2}-4)} \left( \frac{\mu^{2}}{\tau} \right)^{\varepsilon/2}  \\
3 \ \raisebox{-0.48cm}{\includegraphics[height=1.2cm]{psppp_2.pdf}} &= F_{\pp,\qq,\rr} \frac{(d+8)}{64} \frac{(\zeta - 2\lambda )^2}{2-d} \left( \frac{\mu^{2}}{\tau} \right)^{\varepsilon/2} \\
6 \ \raisebox{-0.48cm}{\includegraphics[height=1.2cm]{psppp_3.pdf}}  &= - 3 \ \raisebox{-0.48cm}{\includegraphics[height=1.2cm]{psppp_2.pdf}}
\end{align}
where 
\begin{align}
F_{\pp,\qq,\rr} =&\ \tfrac{1}{3} D\mu^{\varepsilon} \eta \Big[ p^2 (\qq.\rr) + (\pp.\rr) \left(q^2 + 2 (\qq.\rr)\right) \nonumber \\ 
&\ \hspace{1cm} + (\pp.\qq) \left(2 (\pp.\rr)+2 (\qq.\rr) + r^2\right)\Big].
\end{align}
Note that the last two diagrams cancel out similarly to the case of three-point diagrams, and the only non-zero contribution comes only from the classical one-loop 'Model B' diagram. We believe that this is just an artefact of the one-loop approximation as in the case of three-point vertex function. All other diagrams are UV finite, as the external $ \phi' $ field is attached to a three-point vertex (see Sec. \ref{S3:B}). We list them for the completeness
\begin{align}
3 \ \raisebox{-0.48cm}{\includegraphics[height=1.23cm]{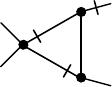}} = 3 \ \raisebox{-0.48cm}{\includegraphics[height=1.23cm]{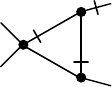}} = 3 \ \raisebox{-0.48cm}{\includegraphics[height=1.23cm]{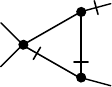}} \nonumber \\
= 6 \ \raisebox{-0.48cm}{\includegraphics[height=1.23cm]{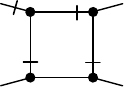}} = 6 \ \raisebox{-0.48cm}{\includegraphics[height=1.23cm]{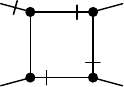}} = 0. \label{APP:eq.43Diag}
\end{align}

\section{Cylindrical coordinates}     \label{APP:CC}

I order to better understand the structure of the RG flow for $ \eta \neq 0 $, we transform the RG flow equations \eqref{S4:eq.RGEq_1}-\eqref{S4:eq.RGEq_2} into cylindrical coordinates
\begin{align}
\lambda &= \rho \cos \theta, \quad \rho = \sqrt{\lambda^{2} + \zeta^{2}}, \label{APP:cc.1} \\
\zeta &= \rho \sin \theta, \quad \theta = \arccos (\lambda/\sqrt{\lambda^{2} + \zeta^{2}} ), \label{APP:cc.2}
\end{align}
with the same $ \eta $. The corresponding beta functions are found using the following relations
\begin{align}
\D_{\mu} \rho &= \frac{\lambda \D_{\mu} \lambda + \zeta \D_{\mu} \zeta }{\sqrt{\lambda^{2} + \zeta^{2}}}, \quad \D_{\mu} \theta = \frac{\lambda \D_{\mu} \zeta + \zeta \D_{\mu} \lambda}{\lambda^{2} + \zeta^{2}}.
\end{align}
which gives
\begin{align}
\beta_{\rho} =&\ - \rho \big( \tfrac{1}{2}\varepsilon + \tfrac{3}{8} \rho^{2} \cos \theta ( 2\sin\theta - \cos\theta) \nonumber \\
&\ \hspace{1cm} + \tfrac{1}{24} \eta (2 \sin (2 \theta )+\cos (2 \theta )-13) \big), \label{S4:eq.Brho} \\
\beta_{\theta} =&\ -\tfrac{1}{12} \eta \rho^{2} \cos\theta ( 2\cos \theta - \sin\theta ), \label{S4:eq.Btheta} \\
\beta_{\eta} =&\ -\eta  \left(\varepsilon - \tfrac{3}{8} \eta + \rho^{2} \cos \theta (\sin \theta - \tfrac{1}{2} \cos \theta ) \right). \label{S4:eq.Beta}
\end{align}
We immediately see from \eqref{S4:eq.Btheta}, that there are two fixed "planes": $ \cos\theta^{*} = 0 $ and $ \tan \theta^{*} = 2 $, which correspond to plane $ \lambda = 0 $, and $ \zeta = 2\lambda $. As $ \beta_{\theta} $ is negative for
\begin{align}
\theta  \in (-\pi/2,\arctan2) \cup (\pi/2,\pi+\arctan2),
\end{align}
and positive otherwise, we conclude that $ \zeta = 2 \lambda $ is a stable plane, while $ \lambda = 0 $ is unstable. The fixed points with $ \lambda^{*} = 0 $ are either FP0 or FPII. For $ \zeta = 2 \lambda $ Eqs. \eqref{S4:eq.Brho}-\eqref{S4:eq.Beta} show three fixed points

\begin{itemize}
	\item FPcI: \hspace{0.4cm} $ \eta ^{*} = \tfrac{4}{3} \epsilon, \quad \rho^{*} = 0 $
	
	\item FPcII: \hspace{0.25cm} $ \eta^{*} = 0, \quad \rho^{*}{}^{2} = -\tfrac{20}{9} \epsilon, \quad \text{for} \ \epsilon < 0, $
	
	\item FPcIII: \hspace{0.1cm} $ \eta^{*} = 4 \epsilon, \quad \rho^{*}{}^{2} = \tfrac{20}{3} \epsilon, \quad \text{for} \ \epsilon > 0 $
	
\end{itemize}
which are exactly FPII, FPI and FPIII$ ^{\pm} $, respectively. 

  \bibliographystyle{apsrev}
  \bibliography{CKPZ.bib} 

\end{document}